\PassOptionsToPackage{square,comma,numbers,sort&compress}{natbib}
\documentclass[Letter,12pt]{elsarticle}
\makeatletter
\long\def\pprintMaketitle{\clearpage
  \iflongmktitle\if@twocolumn\let\columnwidth=\textwidth\fi\fi
  \resetTitleCounters
  \def\baselinestretch{1}%
  \printFirstPageNotes
  \begin{center}%
 \thispagestyle{pprintTitle}%
 \def\baselinestretch{1}%
    \Large\@title\par\vskip18pt 
    \normalsize\elsauthors\par\vskip10pt
    \footnotesize\itshape\elsaddress\par\vskip10pt
    \ifvoid\absbox\else\unvbox\absbox\par\vskip10pt\fi
    \ifvoid\keybox\else\unvbox\keybox\par\vskip10pt\fi
    \end{center}%
  \gdef\thefootnote{\arabic{footnote}}%
  }
\makeatother
\usepackage[utf8]{inputenc}
\usepackage{natbib} 
\usepackage{url} 
\usepackage{verbatim} 
\usepackage{hyperref} 
\hypersetup{
    colorlinks=true,
    linkcolor=blue,
    filecolor=magenta,      
    urlcolor=cyan,
}
\usepackage[left=1in,right=1in,top=1.3in,bottom=1.3in]{geometry}
\usepackage{fancyhdr} 
\usepackage{psfrag}
\usepackage{colortbl}
\usepackage{caption}
\usepackage{amsmath}
\usepackage{physics}
\usepackage{amsfonts}
\usepackage{multirow}
\usepackage{tabularx}
\usepackage{setspace} 
\usepackage[displaymath, mathlines]{lineno} 
\newcommand*\patchAmsMathEnvironmentForLineno[1]{%
  \expandafter\let\csname old#1\expandafter\endcsname\csname #1\endcsname
  \expandafter\let\csname oldend#1\expandafter\endcsname\csname end#1\endcsname
  \renewenvironment{#1}%
     {\linenomath\csname old#1\endcsname}%
     {\csname oldend#1\endcsname\endlinenomath}}%
\newcommand*\patchBothAmsMathEnvironmentsForLineno[1]{%
  \patchAmsMathEnvironmentForLineno{#1}%
  \patchAmsMathEnvironmentForLineno{#1*}}%
\AtBeginDocument{%
\patchBothAmsMathEnvironmentsForLineno{equation}%
\patchBothAmsMathEnvironmentsForLineno{align}%
\patchBothAmsMathEnvironmentsForLineno{flalign}%
\patchBothAmsMathEnvironmentsForLineno{alignat}%
\patchBothAmsMathEnvironmentsForLineno{gather}%
\patchBothAmsMathEnvironmentsForLineno{multline}%
}
\usepackage{subfig}
\usepackage[para,online,flushleft]{threeparttable}

\usepackage{bm}
\usepackage{comment}
\usepackage{xcolor}

\usepackage{ar}
\usepackage{multicol}
\usepackage[export]{adjustbox}
\usepackage{amsthm}
\usepackage{empheq}
\usepackage{gensymb}
\usepackage{graphicx}
\usepackage{enumerate}
\usepackage{enumitem}
\usepackage{esint}
\usepackage{amssymb}
\usepackage{titlesec}
\usepackage{marvosym}
\usepackage{array}
\journal{Journal of Wind Engineering \& Industrial Aerodynamics}
\date{August 19, 2021}
\usepackage{microtype}
\usepackage{soul}
\counterwithout{footnote}{section}

\begin{document}
\begin{frontmatter}

\title{A survey of two analytical wake models for crosswind kite power systems}
\author[1]{Mher M. Karakouzian}
\author[1]{Mojtaba Kheiri\corref{cor1}}
\cortext[cor1]{Corresponding author. Tel.: +1 514 848 2424 ext. 4210. Email: mojtaba.kheiri@concordia.ca}
\address[1]{Fluid-Structure Interactions \& Aeroelasticity Laboratory, Department of Mechanical, Industrial and Aerospace Engineering, Concordia University, 1455 de Maisonneuve Blvd. West, Montr\'{e}al, Qu\'{e}bec H3G 1M8, Canada}
\begin{abstract}
This paper presents two novel analytical wake models for crosswind kite power systems. One is developed based on the continuity equation, and the other based on both the continuity and momentum equations. For each model, equations for the wake flow speed as well as the wake shape are obtained through a rigorous theoretical approach. Wake models for crosswind kites provide a first step in the understanding of the effects of kite-to-kite aerodynamic interactions on prospective wind energy kite farms. Despite a fair number of computational studies on these wakes, few studies have attempted to provide a concrete analytical solution. Indeed, the primary motivation of this study comes from the fact that analytical models are simple in form and require very little computational power when being solved. 
The results from the two analytical models are compared with each other and are validated against computational results. Bearing several assumptions, these models are meant to simply offer a preliminary insight that will hopefully see many improvements with added complexity in the near future.
\end{abstract}
\end{frontmatter}

\section{Introduction}
\label{sec:introduction}
\begin{figure}[!t]
\centering
\includegraphics[scale=.8]{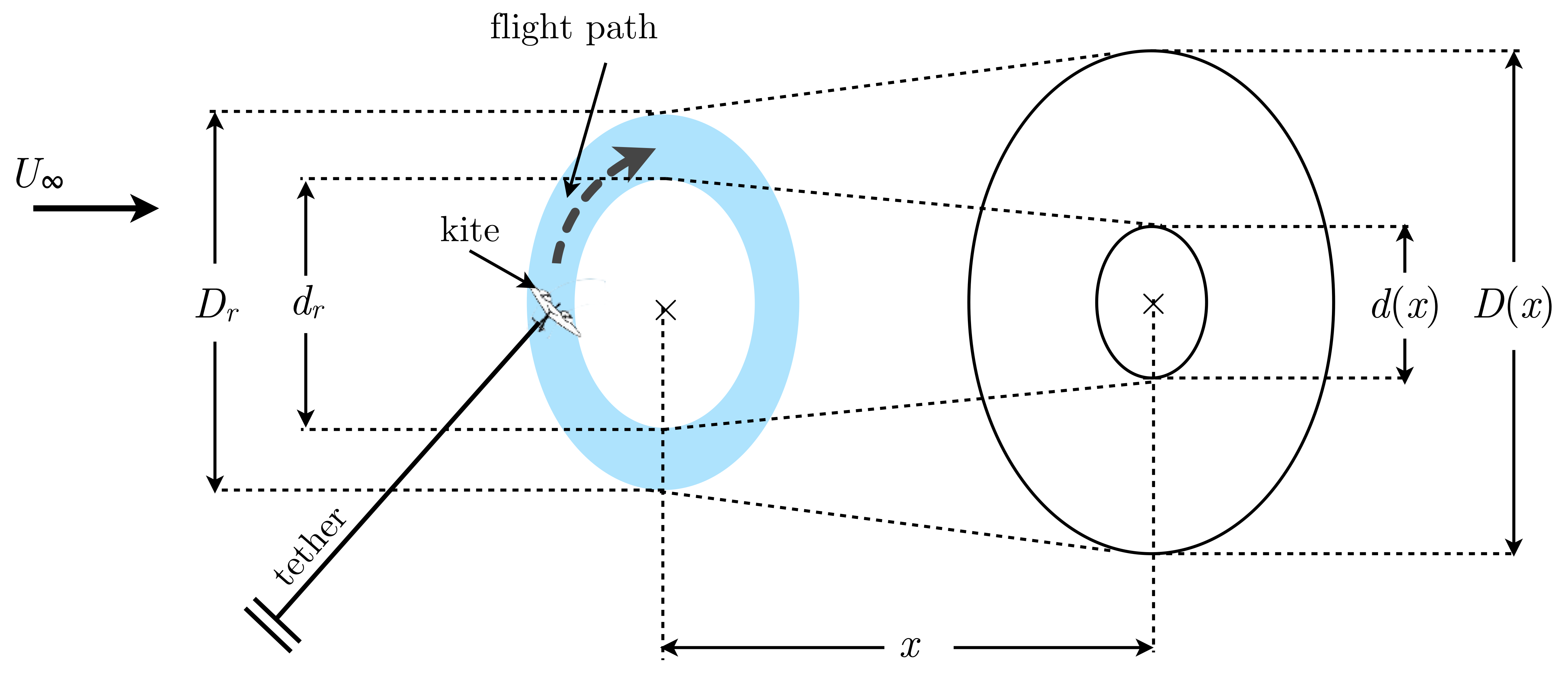}
\caption{General overview of the CKPS, showing the flight path modelled as an annular actuator disc with inner and outer diameters labelled $d_r$ and $D_r$, respectively. An expanding wake is also shown, as a function of downstream distance, $x$, with inner and outer diameters are labelled $d(x)$ and $D(x)$, respectively. Here, $U_\infty$ represents the freestream velocity.}
\label{fig0}
\end{figure}

Airborne wind energy (AWE) is a new and emerging subject in the study of wind energy, garnering huge popularity as of late. Dubbed a \textquotedblleft{}radical innovation breakthrough for the future\textquotedblright{} by the European Commission \cite{EuropianUnion2019}, it involves taking advantage of high-speed winds occurring at high altitudes unattainable by conventional horizontal-axis wind turbines. Because of this, AWE concepts provide a higher power output than conventional wind turbines, while also being a low-cost alternative, as they operate over a tether-and-spool system, rather than being mounted over large and heavy towers. AWE can take many forms, one of which is crosswind kite power systems (CKPSs), the topic of the current study. 
\\ \\
Harnessing power, on a commercial scale, from crosswind motion using kites (typically soft fabric-based or rigid airplane-like structures) is a recent pursuit. Although the idea was originally proposed by Miles L. Loyd \cite{loyd} in 1980. Inspired by his paper, the current-day CKPS concept consists of a kite flying at a high crosswind speed on a circular, figure-eight or spiral trajectory by virtue of oncoming winds, while simultaneously being tethered to a spool on the ground. As such, the kite exploits large aerodynamic forces, while the power is being generated either on the ground by unrolling the tether from a drum, coupled to an electrical generator, or on the kite itself via on-board turbines.
\\ \\
The idea of static kites, namely those which do not employ this crosswind principle, has also been proposed. These, along with several other AWE concepts, have been studied previously by Fagiano and Milanese \cite{fagiano-milanese}, Ahrens et al. \cite{ahrens}, Cherubini et al. \cite{cherubini}, and Schmehl \cite{schmehl}, among others. Nevertheless, there is still much work to be done in the realm of CKPSs. Of particular interest is the aerodynamic interactions between two or more flying kites. Indeed, if a number of such kites were to be arranged together on ``kite farms" or ``parks," as standard modern-day wind turbines are, it would be important to know how one kite affects the other to ensure optimal energy extraction. On this front, some progress has been made; however, the studies have been on mainly mechanical (or physical) interactions. For instance, Fagiano \cite{fagiano} and Faggiani and Schmehl \cite{faggiani-schmehl} have devised effective algorithms for optimal kite arrangement, so as to reduce (or eliminate) the likelihood of kite collision or the entanglement of two or more tethers. Despite this success, however, investigations into kite-to-kite influence due to the wakes they produce downstream are still ongoing and are the primary focus of our study. Indeed, on prospective kite farms, the wake of a given kite will naturally influence the aerodynamics of those nearby. In particular, if a downstream kite is too close, it may experience highly turbulent winds of reduced mean speed. Thus, good-quality wake models inform on the proper distances by which to place one kite from the other.
\\ \\
Despite having a multitude of such models for conventional wind turbines, wake models for CKPSs are still few in number. There have been some computational studies, most notable being those of Haas and Meyers \cite{haas-meyers}, Kheiri et al. \cite{kheiriet2017}, Haas et al. \cite{haaset2019}, and Kheiri et al. \cite{kheiriet2019}. As we will see in next sections, these studies have been instrumental in helping to predict how the shape of the wake changes as a function of the downstream distance from the kite. Still lacking in the literature, however, are analytical models for CKPSs. That is, except for a few recent attempts, there is currently no published closed-form equation that can predict the downstream wake shape or wake velocity to any degree of accuracy. Indeed, the simplicity and low computational cost that such analytical models offer cannot be overlooked.
\\ \\
One attempt has been made by Kaufman-Martin et al. \cite{kaufman} based on the entrainment hypothesis \cite{luzzatto}, but this involves solving a system of three ordinary differential equations. Even with the potential promise of high accuracy, such models are still computationally expensive. Most recently, Kheiri et al. \cite{Kheiri2021} developed a simple analytical wake model which parallels Jensen's \cite{jensen} wake model for conventional wind turbines. This model was found to be simple in form, easy to solve and reasonably accurate. 
\\ \\
With the hopes of expanding the domain of analytical models, in the following sections, we develop two such wake models for CKPSs, each based on one of the available conventional wind turbine wake models: one by Jensen \cite{jensen} and the other by Frandsen \cite{frandsen}. Despite being solely based on the continuity equation, Jensen's wake model (later expanded upon by Katic et al. \cite{katic}) best exemplifies the power and usefulness of analytical models, as, since the time of its publication in 1983, it is still referenced (for example, in Refs. \cite{Kaldellis2021,bossuyt2021quantification}) and used in various commercial software, such as WAsP \cite{BartComparisonSodar2006} and WindPRO \cite{ThogersenComparison2005}. On the other hand, the equally-powerful Frandsen model is based on satisfying both the continuity and momentum equations, thereby providing, as we believe, a more ``holistic" description of the wake flow. It is for this reason that we call our Jensen- and Frandsen-based CKPS models the \textit{Continuity Wake (CW) model} and \textit{Continuity-Momentum Wake (CMW) model}, respectively. Finally, we note that the Jensen-based CW model will be a reformulation of Kheiri et al. \cite{Kheiri2021}, whereas the CMW model will be a completely novel development.
\\ \\
The paper is organized as follows. For each model, we first represent the circular path of the kites as an annular actuator disc, which will hereafter be referred to as the \textit{disc} or the \textit{rotor}. Figure \ref{fig0} shows the general schematic that the models will be based on. Next, we apply actuator disc theory and control volume theory to arrive at equations for the velocity and diameter of the wake as a function of the downstream distance. In the CW model (Section \ref{sec:continuity_wake_model}), the control volume encloses the wake without the rotor, whereas both the wake and the rotor are included in the CMW model (Section \ref{sec:continuity_momentum_wake_model}). After obtaining the equations, we produce numerical results for several crosswind kite systems (Section \ref{sec:numerical_results}). These results are then discussed and compared with computational results from the literature. The paper ends with a summary of major findings (Section \ref{sec:discussion}).

\section{Continuity Wake (CW) model}
\label{sec:continuity_wake_model}
\begin{figure}[!t] 
    \centering
	\includegraphics[scale=.8]{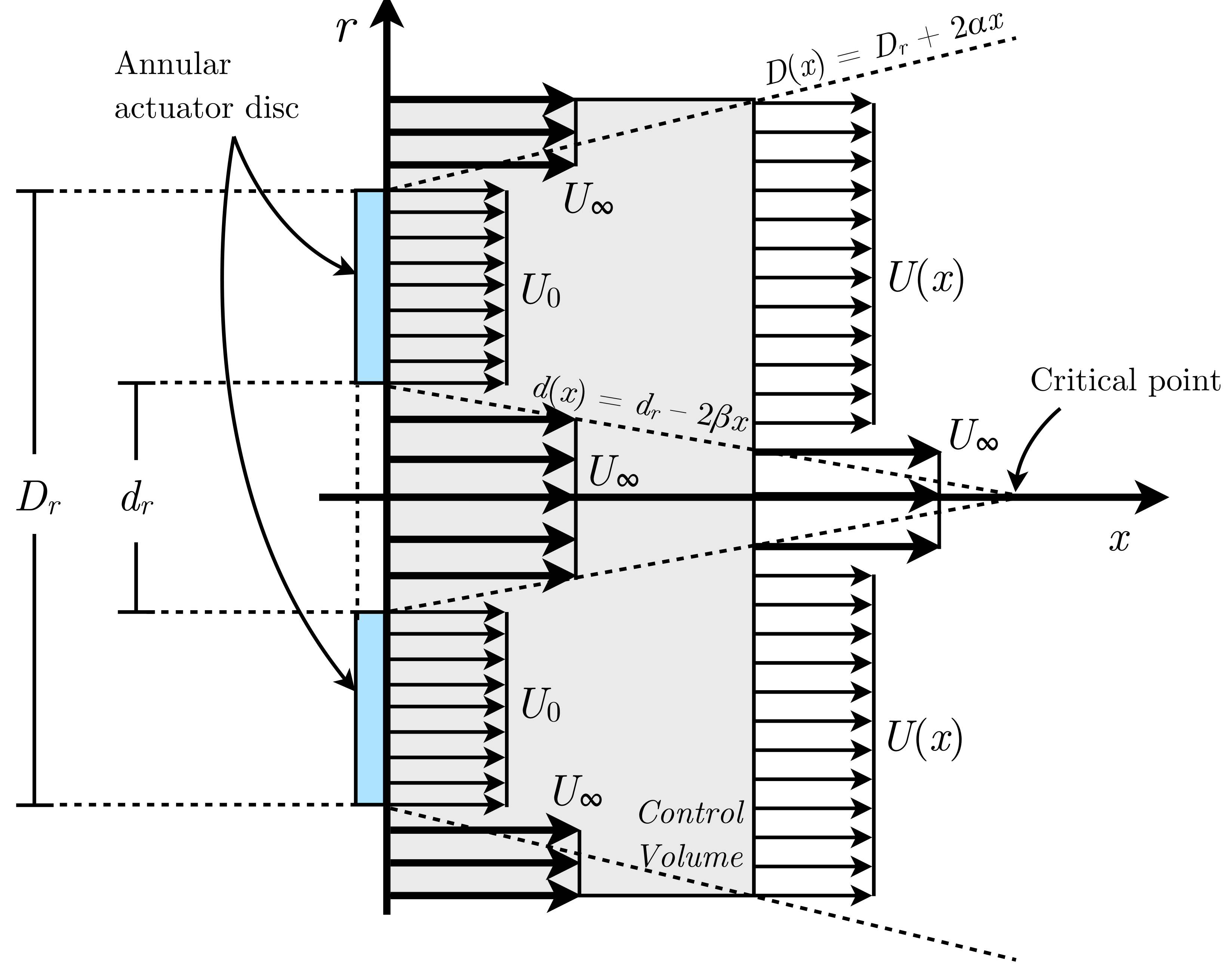}
	\caption{Cross-sectional schematic of the CW model showing the control volume (in grey) starting from the point just 	after the actuator disc/rotor (in blue) and ending at a point downstream, where the wake is fully developed. The ``critical point", where the core region ends, is also shown.}
\label{fig1}
\end{figure}

We start by devising a wake model for the annular actuator disc based only on the continuity equation. This will be a slight reformulation of the derivation found in Kheiri et al \cite{Kheiri2021}. We start by letting $d_r$ and $D_r$ be the inner and outer diameters of the disc, respectively (see Figure \ref{fig0}). We make the following assumptions:

\begin{enumerate}
	\item\label{jasn0} The flow is one-dimensional (or unidirectional) and incompressible.
	\item\label{jasn1} The wake area is distributed linearly over the downstream distance from the rotor (see Figure \ref{fig1}). In this case, we let $2\alpha$ and $2\beta$ be the slopes of the changes in $D(x)$ and $d(x)$, the diameters of the outer and inner portions of the wake, respectively, at a point $x$ downstream. The multiplication by 2 of these constants (sometimes referred to as \textit{entrainment constants}) is because Jensen's model, the model on which the CW model will be based, was derived using radii instead of diameters. This detail will be important later, when we come to perform the numerical analysis. Furthermore, $d$ must eventually become zero, once the core of the annulus integrates itself into the rest of the wake.

		\item\label{jasn2} The flow in the ``core" of the annulus is at the freestream velocity, $U_\infty$, at all times, until it integrates itself with the rest of the wake at some critical point downstream.

		\item\label{jasn3} According to Bernoulli's equation, between the point just behind the rotor and the point downstream where the freestream pressure, $p_\infty$, is recovered (we will refer to this area as the \textit{near-wake region}), there is a velocity deficit. That is, the velocity at the end of this expansion, which is represented by $U_0$, is less than the velocity at the point just after the rotor. For the current model, we assume that the distance over which this expansion occurs is negligibly small relative to the far downstream distances of concern. 
		According to momentum theory, $U_0$ may be written as a fraction of the freestream velocity as follows: $U_0=(1-2a)U_\infty$, $0 \leq a \leq \frac{1}{2}$, where $a$ is called the \textit{induction factor}.

		\item\label{jasn4} The resulting wake flow velocity distribution is uniform; that is, it exhibits a ``top hat" shape.
		\item\label{jasn5} The starting inner and outer diameters of the wake are $d_r$ and $D_r$, respectively. This is consistent with the assumption of negligibly small near-wake region.

	\end{enumerate}

Now, in accordance with Assumptions \ref{jasn0} and \ref{jasn1}, we write
	\begin{equation}\label{eq1}
		D(x) = D_r+2\alpha x,
	\end{equation}
and
	\begin{equation}\label{eq2}
		d(x) =	i_1(x)(d_r-2\beta x),
	\end{equation}
where $i_1(x)=\frac{1}{2}\left(1+\frac{d_r-2\beta x}{|d_r-2\beta x|}\right)$ is an ``integration factor" which ensures that the area of the core eventually goes to zero once flow has integrated itself with the rest of the wake (see Figure \ref{fig1}). 

Now, using $U=U(x)$ for the velocity of the wake, with Assumptions \ref{jasn0} and \ref{jasn2} in mind, we apply continuity:
\begin{equation}\label{eq2-1}
\frac{\pi}{4}(D_r^2-d_r^2)U_0+\frac{\pi}{4}(D^2-D_r^2)U_\infty+\frac{\pi}{4}d_r^2U_\infty=\frac{\pi}{4}(D^2-d^2)U+\frac{\pi}{4}d^2U_\infty.
\end{equation}

\noindent Using Assumption \ref{jasn3}, we can can simplify equation (\ref{eq2-1}) to 

\begin{equation}\label{eq3}
	\frac{U(x)}{U_\infty}=1-2a\frac{D_r^2-d_r^2}{D(x)^2-d(x)^2}.
\end{equation}

\noindent We can non-dimensionalize equations (\ref{eq1}), (\ref{eq2}), and (\ref{eq3}) to get

\begin{equation}\label{eq6-6}
   \begin{aligned}
   \widetilde{D}(\xi) &= \widetilde{D}_r+2\alpha \xi, \\
   \widetilde{d}(\xi) &=i_1(\xi)\left(\widetilde{d}_r-2\beta \xi\right), \\   
    \widetilde{U}(\xi)&=1-2a\frac{\widetilde{D}_r^2-\widetilde{d}_r^2}{\widetilde{D}(\xi)^2-\widetilde{d}(\xi)^2},\;\;\;\;\; 0 \leq a \leq \frac{1}{2} 
   \end{aligned}
\end{equation}
where $\widetilde{U}=U/U_\infty$, $\xi=x/R$, $\widetilde{D}=D/R$, $\widetilde{d}=d/R$, $\widetilde{D}_r=D_r/R$, and $\widetilde{d}_r=d_r/R$, for $R=(D_r+d_r)/4$ the radius of gyration. 

With all the terms substituted, $\widetilde{U}(\xi)$ may be re-written as
\begin{equation}\label{eq6-5}
	\widetilde{U}(\xi)=1-2a\frac{\widetilde{D}_r^2-\widetilde{d}_r^2}{(\widetilde{D}_r+2\alpha\xi)^2-i_1^2(\xi)			(\widetilde{d}_r-2\beta\xi)^2}.
\end{equation}

For future reference, we also note that the final solution in equation (\ref{eq6-6}) can also be written as

\begin{equation}\label{eq6}
	\widetilde{U}(\xi)=1-2a\frac{1}{\widetilde{A}(\xi)},
\end{equation}

\noindent where $\widetilde{A}=A/A_r$ for $A=A(\xi)=\frac{\pi}{4}[D^2(\xi)-d^2(\xi)]$ and $A_r=\frac{\pi}{4}(D_r^2-d_r^2)$ the areas of the wake and the rotor, respectively.
\\
\section{Continuity-Momentum Wake (CMW) model}
\label{sec:continuity_momentum_wake_model}
\begin{figure}[!t] \centering
	\includegraphics[scale=.8]{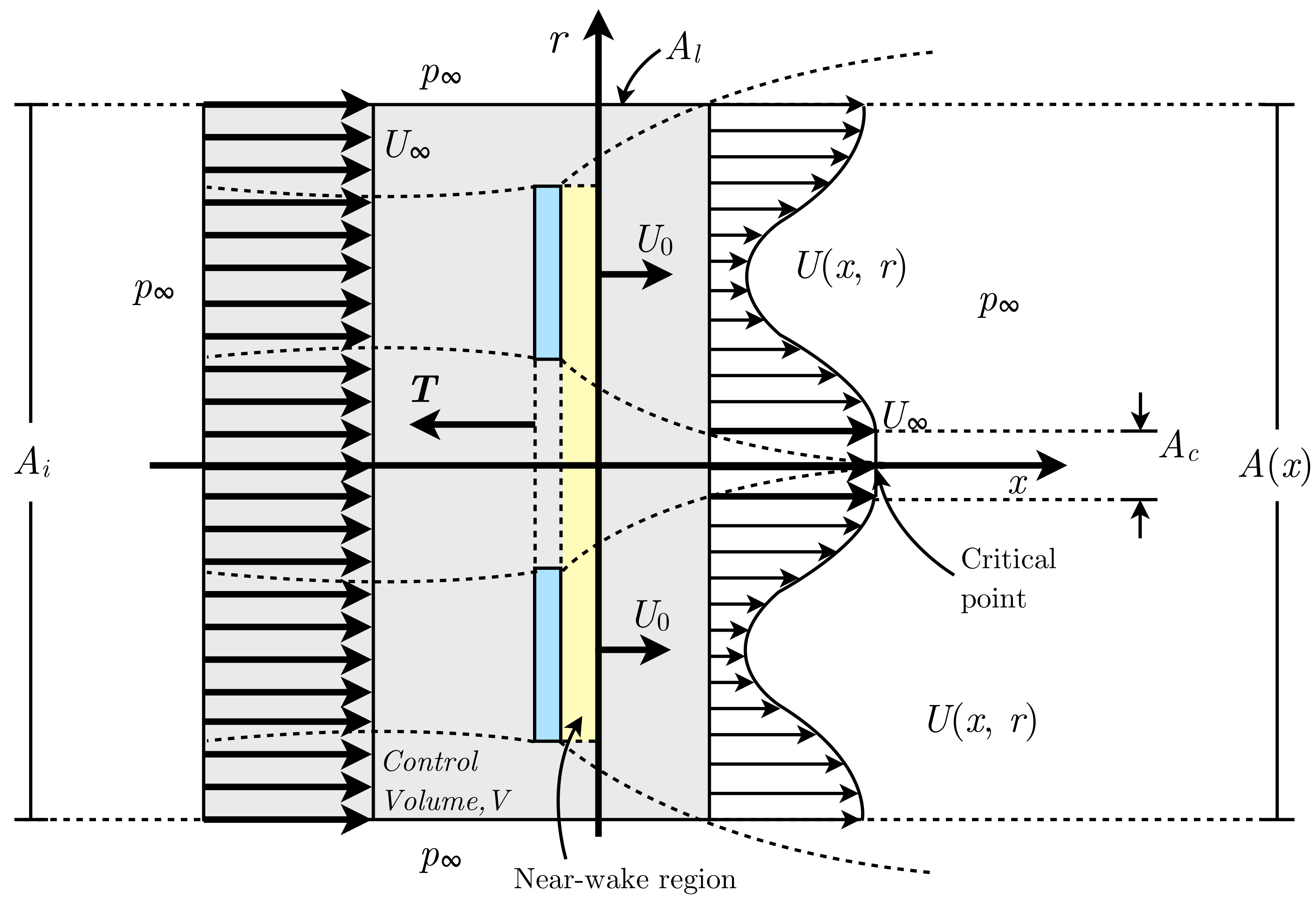}
	\caption{Schematic of the CMW model, showing the control volume (in grey) of volume $V$ and surface area $S$, the thrust, $\boldsymbol{T}$, imposed on the rotor (in blue), and the velocity at the end of the initial expansion, $U_0$. Thus, the pressure around the entire control volume is equal to the freestream pressure, $p_\infty$. Shown also are the near-wake region (in yellow) and the ``critical point", where the core region ends. It can be seen that the resulting flow speed is assumed to be equal to the freestream velocity within the core area, $A_c$ and (a priori) assumed to be arbitrary in the annular region, $A(x)$.}
\label{fig2}
\end{figure}

In this section, we derive an expression for the wake velocity and diameter based on both the continuity and momentum equations. Proceeding as in Frandsen et al. \cite{frandsen}, we first consider a cylindrical control volume of surface area $S$ and of volume $V$ around the whole actuator disc. The wake speed, $U$, will be defined at a point $x$ downstream coinciding with the outlet of this cylinder. Figure \ref{fig2} shows the cross-sectional view.
Here, $\boldsymbol{T}$ is the thrust exerted on the actuator disc (the remaining variables in the figure are defined as in the previous section). Then, starting from the momentum equation, we have
\begin{equation}\label{eq7}
\int_V\rho\pdv{\boldsymbol{U}}{t}\;dV+\int_S\rho\boldsymbol{U}(\boldsymbol{U}\cdot d\boldsymbol{S})=-\int_Sp\;d\boldsymbol{S}+\int_V\rho\boldsymbol{g}\;dV+\boldsymbol{T}+\int_S\boldsymbol{\tau}\;d\boldsymbol{S}.
\end{equation}
We now make the following assumptions:

\begin{enumerate}

	\item\label{fasn1} The flow is inviscid, so that the shear stress tensor is zero: $\boldsymbol{\tau}=\textbf{0}$.

	\item\label{fasn2} Compressibility effects are negligible: $\rho$ is constant.

	\item\label{fasn3} Gravitational effects are negligible: $\boldsymbol{g}=\textbf{0}$.

	\item\label{fasn4} The flow is steady: $\dfrac{\partial{\boldsymbol{U}}}{\partial t}=\textbf{0}$.

	\item\label{fasn5} All faces of the control volume are under the freestream pressure, $p_\infty$: $-\int_Sp\;d\boldsymbol{S}=0$.

	\item\label{fasn6} Any rotation of the flow produced by the rotor is negligible, so that the flow is axisymmetric: $U(x,r,\theta)=U(x,r)$.

	\item\label{fasn7} The magnitude of the velocity is monotonically increasing (i.e. never drops) after the initial expansion.
	
	\item\label{fasn8} The flow in the ``core" of the annulus is at the freestream velocity, $U_\infty$, at all times, until 	it integrates itself with the rest of the wake at some critical point downstream.

\end{enumerate}

\noindent Then, by Assumptions \ref{fasn1} to \ref{fasn5} above, equation (\ref{eq7}), in the axial $x$-direction, reduces to

\begin{equation}\label{eq7-1}
	-T=-\int_{A_i}\rho U_\infty^2\;dA+\int_{A}\rho U^2\;dA+\int_{A_c}\rho U_\infty^2\;dA + \int_{A_l}\rho (\boldsymbol{U} \cdot \boldsymbol{\widehat{e}_x}) (\boldsymbol{U} \cdot d \boldsymbol{S}),
\end{equation}

\noindent where $A_i$, $A$, $A_c$, and $A_l$ are, respectively, the areas of the control volume inlet, annular portion of the outlet (i.e. wake), core portion of the outlet, and lateral/curved surface (see Figure \ref{fig2}); also, $\boldsymbol{\widehat{e}_x}$ is the unit vector in the $x$-direction. Note that, given the way the control volume is selected, the sum of the outlet areas (core and annular) is equal to the inlet area.
\\ \\
As explained in \ref{sec:appendix_A}, after applying the continuity equation, equation (\ref{eq7-1}) may be simplified to

\begin{equation}\label{eq7-5}
	T=\int_A\rho U(U_\infty-U)\;dA.
\end{equation}

\noindent Frandsen et al. \cite{frandsen} showed that an additional assumption of self-similarity of the flow velocity results in a decoupling of the $x$ and $r$ variables. With this, they eventually reasoned that the assumption of $U(x)$ having a rectangular, ``top hat", distribution (as in the CW model) is fair. This makes $U$ independent of $r$, so combined with Assumption \ref{fasn6}, we have $U(x, r)=U(x)$. Thus, due to the decoupling, at any fixed point on the $x$-axis, equation (\ref{eq7-5}) is seen as being

\begin{equation}\label{eq8}
	T=\rho AU(U_\infty-U).
\end{equation}


\noindent Now, we know that, by definition,

\begin{equation}\label{eq8-5}
	T=\frac{1}{2}\rho A_rU_\infty^2C_T,
\end{equation}

\noindent where $A_r$ is defined as before and $C_T$ is the thrust coefficient. Then, equating (\ref{eq8}) and (\ref{eq8-5}), we get

\begin{equation}\label{eq9}
	\frac{U(x)}{U_\infty}=\frac{1}{2}\pm\frac{1}{2}\sqrt{1-2C_T\frac{A_r}{A(x)}}.
\end{equation}

\noindent According to Assumption \ref{fasn7}, the ``$\pm$" sign in this equation must be ``$+$." Indeed, otherwise we would have that ${U(x)}/{U_\infty}\rightarrow 0$ as $x\rightarrow \infty$ (instead of ${U(x)}/{U_\infty}\rightarrow 1$, which is what we wish to have in order to match the physics of the problem). Thus, equation (\ref{eq9}) should be

\begin{equation}\label{eq9-1}
	\frac{U(x)}{U_\infty}=\frac{1}{2}+\frac{1}{2}\sqrt{1-2C_T\frac{A_r}{A(x)}}.
\end{equation}

To decide the point at which to set the origin $x=0$, we turn again to the CW model. An important feature of this model is that the origin $x=0$ is set at the end of the initial wake expansion. This is a natural choice, as the pressure has recovered to the freestream pressure here. However, we must make sure that, with such a choice, equation (\ref{eq9-1}) has solutions for values of $a$ between $0$ and ${1}/{2}$, as per its definition.\footnote{Please note that in the formulation of Frandsen et al. \cite{frandsen}, the variable \textquotedblleft{}$a$\textquotedblright{} was used to represent the induction factor at the end of initial wake expansion (rather than at the rotor), which according to the classical momentum theory is two times \textquotedblleft{}$a$\textquotedblright{} used in this study.} To this end, suppose that, indeed, $U(0)=U_0$ and consider, first, the continuity equation applied to the near-wake region (see Figure \ref{fig2}): $\rho A_rU_r=\rho A_0 U_0$. Then, 
\begin{equation}\label{eq9-2}
	A_0=c A_r,
\end{equation}
where $c=(1-a)/(1-2a)$ is the area ratio. Since also $A(0)=A_0$, substituting $U(0)=U_0$, equation (\ref{eq9-2}), and the identity from momentum theory

\begin{equation}\label{eq8-6}
	C_T=4a(1-a)
\end{equation}

\noindent into equation (\ref{eq9-1}) gives the following:

\begin{equation}\label{eq9-3}
	\frac{U_0}{U_\infty} =	\begin{cases}
							2a & a > \frac{1}{4} \\
							1-2a & a \leq \frac{1}{4}
						\end{cases}.
\end{equation}

\noindent As mentioned, momentum theory dictates that ${U_0}/{U_\infty}=1-2a$. Thus, according to equation (\ref{eq9-3}), equation (\ref{eq9-1}) has solutions only for $0 \leq a \leq{1}/{4}$, which is in the mentioned acceptable range of $0$ and $1/2$. \\ \\

Finally, it is left to determine an appropriate wake area distribution, $A(x)$. We provide a modified form of what is given in Ref. \cite{frandsen}. We start by assuming that the change in wake inner diameter in the near-wake region is negligible; that is

\begin{equation}\label{eq9-4}
	d_0=d_r,
\end{equation}

\noindent where the subscript $0$ denotes the point at the end of the initial expansion. This is a rather rough approximation, but also necessary to apply the methods of Frandsen. Since $A_r=\frac{\pi}{4}(D_r^2-d_r^2)$ and $A_0=\frac{\pi}{4}(D_0^2-d_0^2)$, according to equation (\ref{eq9-2}), this means that we must also have 

\begin{equation}\label{eq9-5}
D_0=\sqrt{c}\sqrt{D_r^2-\left(1-\frac{1}{c}\right)d_r^2},
\end{equation}
where it can easily be shown that $D_0>D_r$. \\ \\

Given these two definitions for the outer and inner diameters for the wake just after the initial expansion, we are able to use a similar approximation found in \cite{frandsen} to define the outer and inner diameters for the general wake.\footnote{These definitions (inspired by jet flow) come from the fact that, according to classical turbulence theory \cite{tennekes2018first}, wake diameter is asymptotically related to the downstream distance: that is, $D,d \sim x^{1/k}$.}\label{foot:k_value} The expressions for the wake velocity and inner and outer diameters in dimensionless form are given as
\begin{equation}\label{eq14-1}
   \begin{aligned}
      \widetilde{U}(\xi)	&=	\frac{1}{2}+\frac{1}{2}\sqrt{1-8a(1-a)\frac{\widetilde{D}_r^2-\widetilde{d}_r^2}{\widetilde{D}(\xi)^2-\widetilde{d}(\xi)^2}},\;\;\;\;\; 0 \leq a \leq \frac{1}{4} \\
      \widetilde{D}(\xi)	&=	(c^{k/2}+\varphi \xi)^{1/k}\sqrt{\widetilde{D}_r^2-\left(1-\frac{1}{c}\right)\widetilde{d}_r^2}, \\
      \widetilde{d}(\xi)	&=	i_2(\xi)(1-\psi \xi)^{1/k}\widetilde{d}_r,
   \end{aligned}
\end{equation}
where $\varphi$ and $\psi$ serve as entrainment constants, $i_2(\xi)=\dfrac{1}{2}\left(1+\dfrac{1-\psi \xi}{|1-\psi \xi|}\right)$ is an integration factor (similar to that of $i_1(\xi)$ in the previous section), and $k$ is a constant which depends on the wake distribution model used. Note also that $\widetilde{U}$, $\widetilde{D}$, $\widetilde{d}$, $\widetilde{D}_r$, $\widetilde{d}_r$, and $\xi$ are defined as in Section \ref{sec:continuity_wake_model}. \\ \\



Before moving on to the numerical simulation, we make one final observation. For $C_T \ll 1$, we may use the first-term Taylor series approximation for $\sqrt{1-x} \approx 1 - x/2 + \mathcal{O}(x^2)$ on equation (\ref{eq9-1}) and rewrite it as

\begin{equation}\label{eq11}
	\frac{U(x)}{U_\infty} \approx 1-\frac{1}{2}\frac{A_r}{A(x)}C_T.
\end{equation}

\noindent Now, from momentum theory, equation (\ref{eq8-6}) may be solved for $a$ to give

\begin{equation}\label{eq10}
	a=\frac{1}{2}-\frac{1}{2}\sqrt{1-C_T}.
\end{equation}

\noindent Using, again, the same Taylor approximation on this equation, for $C_T\ll1$, and applying the non-dimensional variables, equation (\ref{eq11}) becomes

\begin{equation}\label{eq12}
	\widetilde{U}(\xi) \approx 1-2a\frac{1}{\widetilde{A}(\xi)}.
\end{equation}


As Frandsen et al. point out in \cite{frandsen}, the likeness of equations (\ref{eq6}) and (\ref{eq12}) suggests the following simplification: we assume that there is at least one point $\xi_0>0$ downstream, where the diameters of both models coincide. Then, equating the expressions for $\widetilde{D}$ in equations (\ref{eq6-6}) and (\ref{eq14-1}), we get

\begin{equation}\label{eq16}
	(c^{k/2}+\varphi \xi_0)^{1/k}\sqrt{\widetilde{D}_r^2-\left(1-\frac{1}{c}\right)\widetilde{d}					_r^2} =\left(\widetilde{D}_r+2\xi_0\alpha\right),
\end{equation}
which may be re-written as follows to give
\begin{equation}
    \varphi = \dfrac{c^{k/2}}{\xi_0} \left[ \frac{\left(\widetilde{D}_r + 2 \xi_0 \alpha\right)^k}{c^{k/2}\left(\widetilde{D}_r^2 - \left(1 - \dfrac{1}{c}\right) \widetilde{d}_r^2\right)^{k/2}} - 1 \right].
\end{equation}








Applying the above method to the inner diameter, we find a similar expression for $\psi$:
\begin{equation}\label{eq18}
	\psi=\frac{1}{\xi_0}\left[1-\left(1-\frac{2\xi_0 \beta}{\widetilde{d}_r}\right)^k\right].
\end{equation}
\\
\section{Numerical results}
\label{sec:numerical_results}
We now plot the wake flow distribution and wake shape over the downstream distance $\xi$. We show three simulations: one to be compared to the CFD results of Kheiri et al. \cite{kheiriet2019} and two to be compared with the CFD results of Haas and Meyers \cite{haas-meyers} (one each for both their turbulent inflow (TI) and laminar inflow (LI) results). We take $k=2$ (as opposed to $k=3$, which is what is used in classical turbulence theory \cite{tennekes2018first}), since, as pointed out in Frandsen et al. \cite{frandsen}, this ensures a ``non-vanishing and non-increasing flow speed" for turbines (and hence, kites) arranged together on wind parks. In addition, we choose Frandsen et al.'s value of $\xi_0=7$. Other values of $\xi_0$ would also be equally fine, 
as it was found that our results were not sensitive to the choice of $\xi_0$.
\\ \\
We test our models under four different combinations of entrainment constants $\alpha$ and $\beta$ (labelled as Case 1 to 4, below). From studies on turbulent mixing in the planetary boundary layer (see, e.g., \cite{stull1988introduction,frandsen1992wind,pena2014atmospheric}), entrainment constants have been shown to be related to atmospheric conditions (and thus, altitude), properties of the air flow, and terrain roughness. Indeed, Frandsen \cite{frandsen1992wind} has provided the following semi-empirical formula for wind turbines:
\begin{equation}
\alpha=\frac{1}{2\ln(h/z_0)},  
\end{equation}
where $h$ is the turbine hub-height and $z_0$ is the surface roughness of the terrain. This formula has been shown to be quite reliable, as many of the entrainment constants used in the literature have been shown to obey it. Indeed, for conventional wind turbines, this formula yields a range of 0.04 to 0.06 (here, typical values for $h$ and $z_0$ have been taken from \cite{henderson1993tropical} and \cite{wieringa1998far}, respectively), and in the literature, the recommended values are between 0.04 and 0.05 for offshore turbines (e.g., see \cite{barthelmie2010evaluation,cleve2009model,barthelmie2006comparison,barthelmie2007modelling}) whereas generally higher values such as 0.075 and 0.1 (e.g., see \cite{barthelmie2006comparison,Bastankhah2014}) are used for onshore turbines. Although, to the best of our knowledge, there are no published analytical CKPS wake models that employ the use of entrainment constants, computing as before for conventional wind turbines, the above formula yields values around 0.05 for CKPSs. On the other hand, since values as high as 0.1 have previously been used in the literature for conventional wind turbines, 
in this study, we have chosen to examine entrainment constant pairs ($\alpha$ and $\beta$) of 0.05 and 0.1. However, we note that we have chosen smaller values for the Haas and Meyers laminar inflow (LI) CFD results (see Simulation III below), since, due to lack of momentum transfer, we expect the wake to expand slower under laminar flow conditions. These smaller values were simply 0.05 and 0.1 divided by two, since this ensured the wake expansion was reasonably gradual and that these values ended up within a fair range of the entrainment constants obtained by curve-fitting to CFD results.


\subsection{Simulation I: Comparison with Kheiri et al. results}
\label{subsec:simulation_I_comp}
\begin{figure}[!h]
    \centering
    \subfloat[\label{fig:comp_simulation_I_radius_0.1_0.1}]{\includegraphics[scale=0.24]{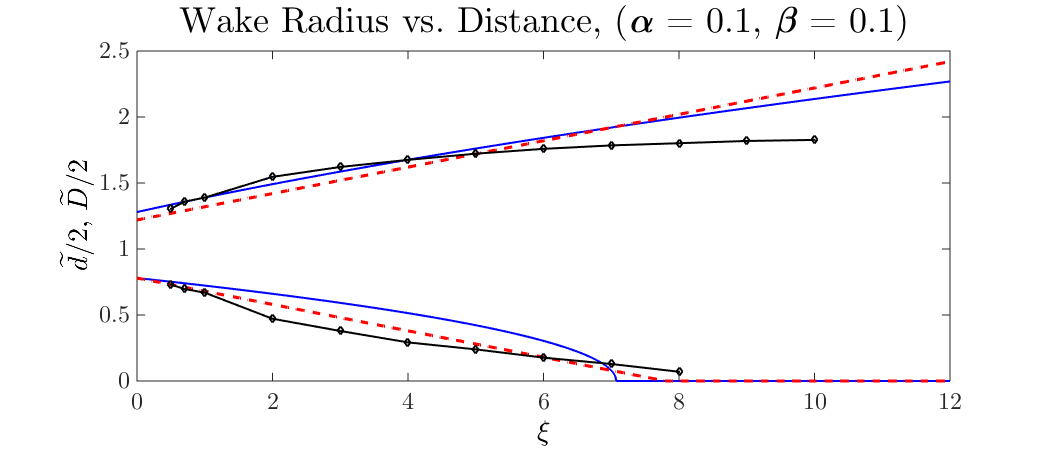}}
    \subfloat[\label{fig:comp_simulation_I_velocity_0.1_0.1}]{\includegraphics[scale=0.24]{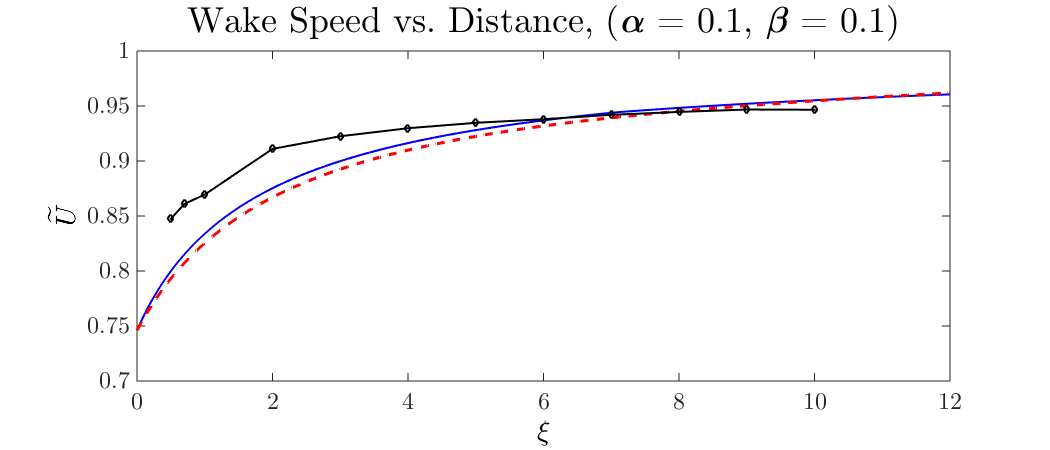}}\\
    \subfloat[\label{fig:comp_simulation_I_radius_0.05_0.05}]{\includegraphics[scale=0.24]{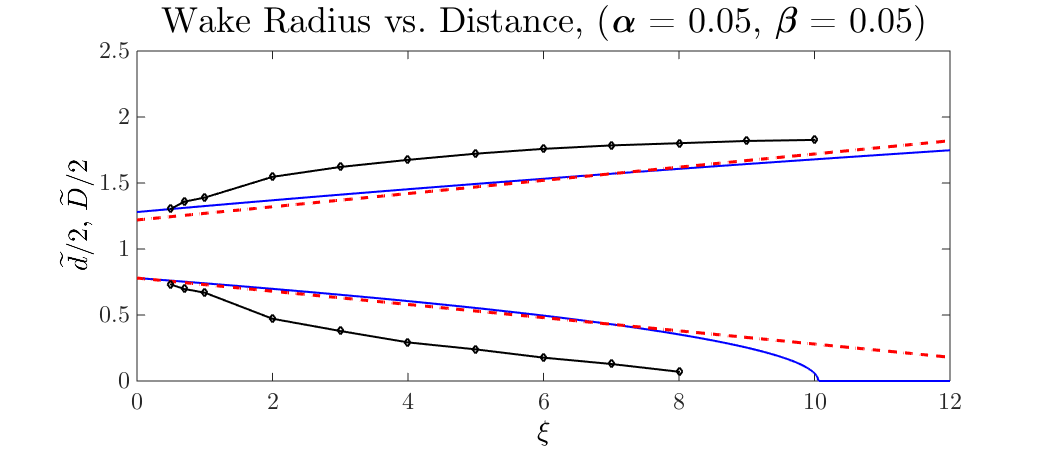}}
    \subfloat[\label{fig:comp_simulation_I_velocity_0.05_0.05}]{\includegraphics[scale=0.24]{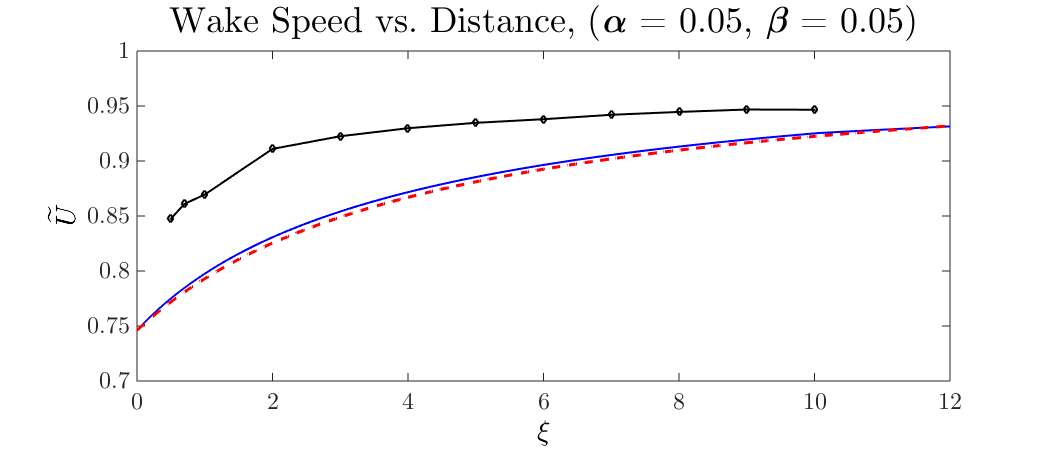}}\\  
    \subfloat[\label{fig:comp_simulation_I_radius_0.05_0.1}]{\includegraphics[scale=0.24]{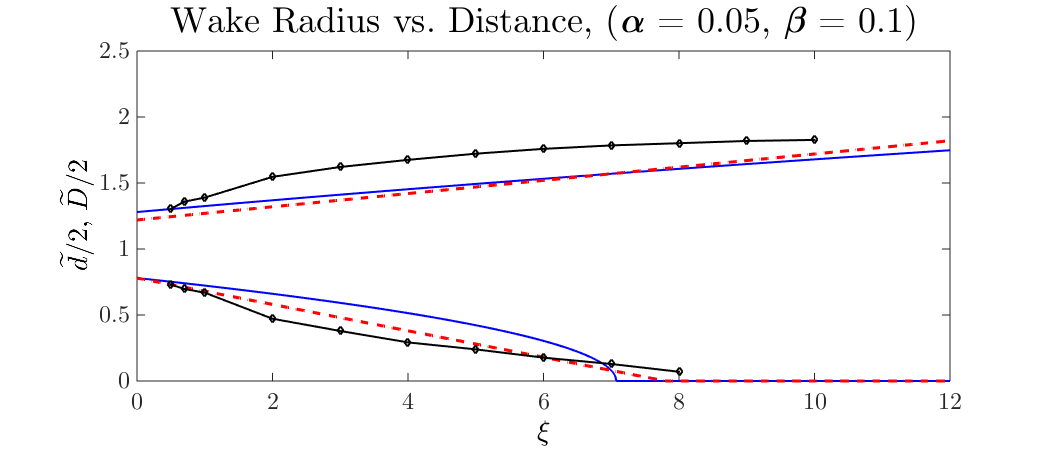}}
    \subfloat[\label{fig:comp_simulation_I_velocity_0.05_0.1}]{\includegraphics[scale=0.24]{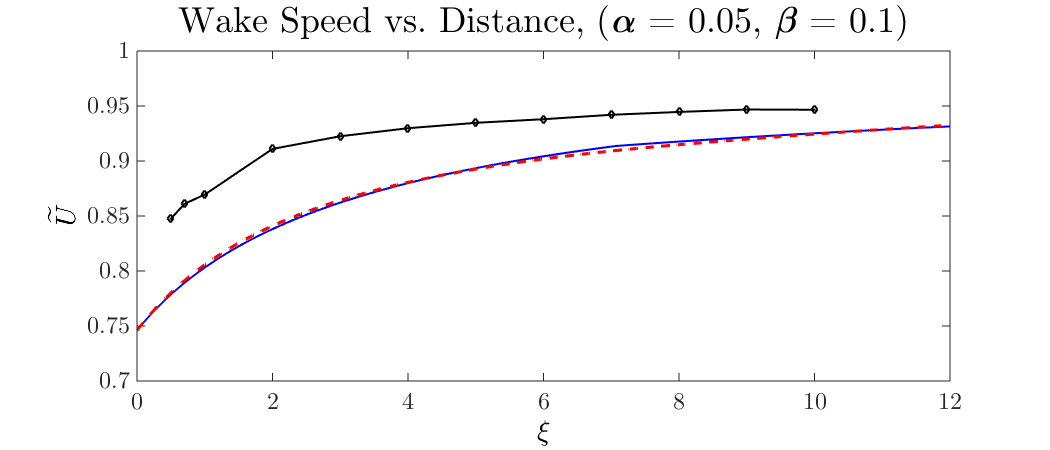}}\\
    \subfloat[\label{fig:comp_simulation_I_radius_0.058_0.091}]{\includegraphics[scale=0.24]{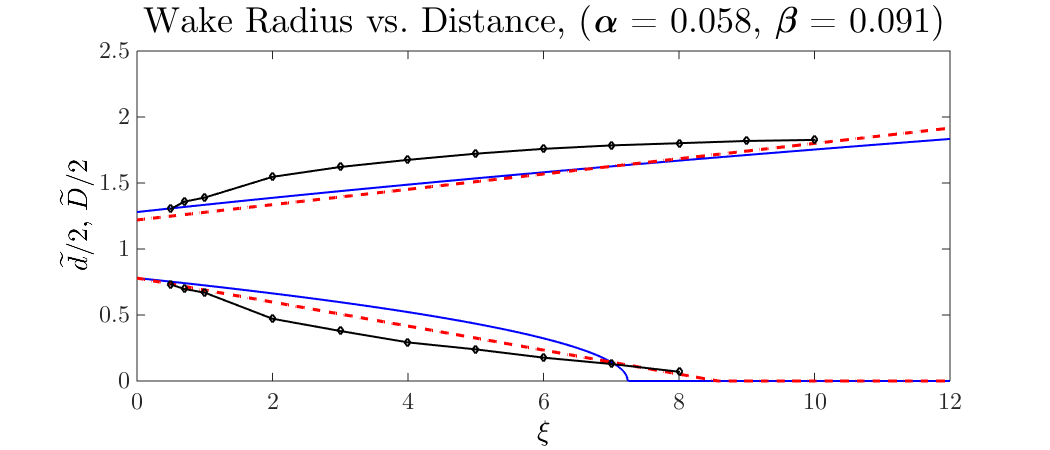}}
    \subfloat[\label{fig:comp_simulation_I_velocity_0.058_0.091}]{\includegraphics[scale=0.24]{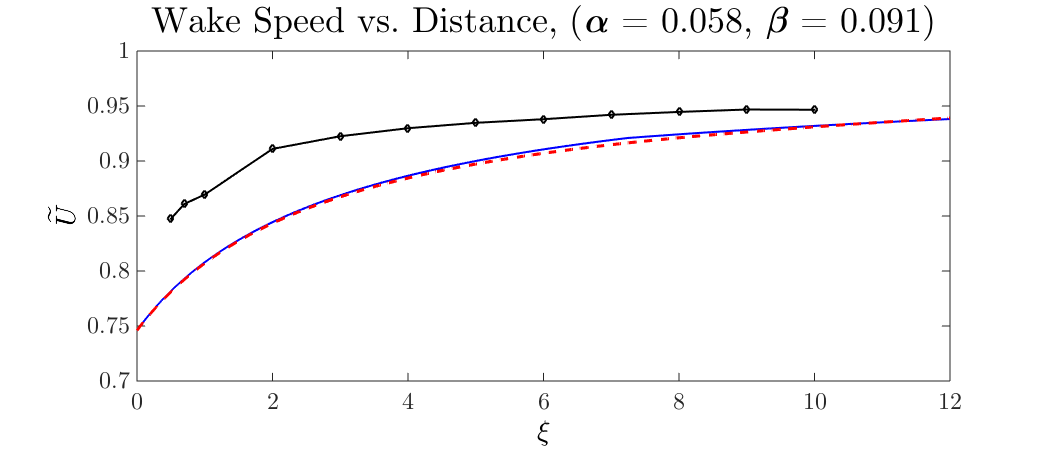}}   
    \caption{Variation of (\textit{left}) dimensionless wake radii and (\textit{right}) dimensionless wake velocity versus dimensionless distance downstream of the rotor for the CW model (dashed line), CMW model (solid line), and CFD model of Kheiri et al. \cite{kheiriet2019} (line with markers). Various entrainment constant pairs were used in the CW and CMW models: (a,b) $\alpha=0.1$, $\beta=0.1$, (c,d) $\alpha=0.05$, $\beta=0.05$, (e,f) $\alpha=0.05$, $\beta=0.1$, and (g,h) $\alpha=0.058$, $\beta=0.091$.}
    \label{fig:comp_simulation_I}
\end{figure}

The first comparison of the theoretical results was made with the CFD results of Kheiri et al. \cite{kheiriet2019}. In this study, the kite (a wing with fixed dimensions) was assumed to be flying at a constant angular velocity, $\Omega=0.738$ rad/s, in a circular path of radius $R=123.3$ m, against an oncoming wind speed of $U_\infty=8.33$ m/s (assumed to be spatially uniform and flowing normal to the plane of rotation) with a turbulence intensity of 1\%. The complete set of parameters for the kite setup are summarized at the top of Table \ref{tab:key_parameters_Kheiri_et_al}, where $b$ and $A_k$ are the kite/wing's span and area, respectively, $Re$ is the Reynolds number of the relative flow, and $D_r$, and $d_r$ are defined as before. Note that these values give the dimensionless diameters as $\widetilde{D}_r=2.44$ and $\widetilde{d}_r=1.56$. The CFD solution domain was made very large (relative to $R$) to minimize the flow blockage. Moreover, the plane of rotation was placed 4.5 km from the inlet and 15.5 km from the outlet to prevent interactions between the domain boundaries and the wake flow. For more details, the interested reader is referred to \cite{Kheiri2018} which used a similar CFD simulation approach for the same kite system as above.
The bottom half of Table \ref{tab:key_parameters_Kheiri_et_al} shows the details of the CFD parameters.
\\ \\
To compare with the analytical models, derived above, we take four test cases for $\alpha$ and $\beta$ as given in Table \ref{tab:test_cases_simulation_I}. As seen from the table, for the first three cases, the entrainment constants $\alpha$ and $\beta$ are different combinations of 0.05 and 0.1; those for Case 4, i.e. $\alpha=0.058$ and $\beta=0.091$, on the other hand, have been obtained by fitting lines to the wake width obtained from the CFD simulation. As seen, a larger entrainment constant has been obtained for the wake flow expansion towards the core region. In other words, the wake inner radius decreases at a higher rate compared to the rate of increase of the outer radius. Moreover, the induction factor was computed by inspection of the CFD flow velocity contour on the plane of rotation and was found to be $a=0.127$.
\\ \\
Figure \ref{fig:comp_simulation_I} shows a side-by-side comparison of (\textit{left}) the wake shape/radius and (\textit{right}) wake speed for Simulation I, where the CW and CMW models are represented by the dashed red line and solid blue line, respectively, and the CFD model of Kheiri et al. \cite{kheiriet2019} is represented by the line with the markers. Since the wake was assumed to be symmetric, only half of the wake cross-section is shown in this figure and those following. Looking at the figure, we can see that, due to the linear (or near-linear) nature of the models, there is large discrepancy between the expansion of the wake radii found computationally and analytically. The CFD wake radii appear as almost bi-linear curves as a function of $\xi$, possibly because of different flow dynamics occurring near and far from the kite; this difference between near wake and far wake flow dynamics is well-known for conventional wind turbines; please refer to \cite{crespo1999survey} for details. However, the curves for the radii for the two analytical models are in good agreement with each other and the model using the curve-fitted entrainment constants seems to predict the critical point (i.e. where $d$ vanishes) quite well. Looking at the wake speed graphs, we see that the CFD curve reaches a plateau much quicker than the curves of the analytical models --- the wake velocity changes only slightly beyond $\xi=4$. In most cases presented in the figure, the discrepancy between CFD and analytical wake velocities decreases monotonically. For example, as shown in Figure \ref{fig:comp_simulation_I} \subref{fig:comp_simulation_I_velocity_0.05_0.1}, the relative error decreases from 8\% at $\xi=0.5$ to 4\% at $\xi=5$ and down to 2\% at $\xi=10$. Lastly, the wake speed curves of the two analytical models align almost perfectly. This could be because, due to the nature of how the diameters were modelled (see equations (\ref{eq6-6}) and (\ref{eq14-1})), a low induction factor, like the one for the kite system in Kheiri et al.'s work, brings the CMW model closer mathematically to the CW model.
\begin{table}[!t]
    \centering
    \caption{Key parameters of the kite system of Kheiri et al. \cite{kheiriet2019} and their CFD simulation set-up.}
    \label{tab:key_parameters_Kheiri_et_al}
    \resizebox{\textwidth}{!}{
    \begin{tabular}{|c|c|c|c|c|c|c|c|c|}
       \hline
       \multicolumn{9}{|c|}{\cellcolor[gray]{0.9}\textbf{Kite System Parameters}}\\
       \hline
       $\boldsymbol{b\textbf{\;(m)}}$  & $\boldsymbol{A_k}$ \textbf{(m}$\boldsymbol{{}^2}$\textbf{)} & $\boldsymbol{R}$ \textbf{(m)} & $\boldsymbol{D_r}$ \textbf{(m)} & $\boldsymbol{d_r}$ \textbf{(m)} & $\boldsymbol{\Omega}$ \textbf{(rad/s)} & $\boldsymbol{U_\infty}$ \textbf{(m/s)} & $\boldsymbol{Re}$ & \textbf{Airfoil}
       \\
       \hline
        53.94 & 200.7 & 123.3 & 300.54 & 192.66 & 0.738 & 8.33 & $20 \times 10^6$ & Clark Y\\
        \hline \hline
        \multicolumn{9}{|c|}{\cellcolor[gray]{0.9} \textbf{CFD Parameters}}\\
        \hline
        \multicolumn{3}{|c|}{\textbf{Domain size (W}$\boldsymbol{\times}$\textbf{H}$\boldsymbol{\times}$\textbf{L)} \textbf{(km)}} & \textbf{Blockage ratio (\%)} & \multicolumn{2}{|c|}{\textbf{Grid resolution/count}} & \multicolumn{2}{|c|}{\textbf{Solver type}} & \textbf{Software} \\
        \hline
        \multicolumn{3}{|c|}{$9\times 9 \times 20$} & 0.05 & \multicolumn{2}{|c|}{6.9 millions} & \multicolumn{2}{|c|}{URANS+$k-\omega$ SST} & Ansys Fluent \\
        \hline
    \end{tabular}}
\end{table}
\begin{table}[!t]
    \centering
    \caption{Test Cases for Simulation I}
    \label{tab:test_cases_simulation_I}
    \begin{tabular}{|c|c|c|}
       \hline
       \cellcolor[gray]{0.9} & \multicolumn{2}{|c|}{\cellcolor[gray]{0.9}\textbf{Entrainment Constant}} \\
       \cline{2-3}
        \multirow{-2}{*}{\cellcolor[gray]{0.9}\textbf{Case Number}}  & $\boldsymbol{\alpha}$ & $\boldsymbol{\beta}$ \\
        \hline
        \cellcolor[gray]{0.9} \textbf{1} & 0.1 & 0.1 \\
        \hline
        \cellcolor[gray]{0.9} \textbf{2} & 0.05 & 0.05 \\
        \hline
        \cellcolor[gray]{0.9} \textbf{3} & 0.05 & 0.1 \\
        \hline
        \cellcolor[gray]{0.9} \textbf{4} & 0.058 & 0.091 \\
        \hline
    \end{tabular}
\end{table}
\clearpage
\subsection{Simulation II: Comparison with Haas and Meyers Results for Turbulent Inflow}
\label{subsec:simulation_II_comp}
\begin{figure}[!t]
    \centering
    \subfloat[\label{fig:comp_simulation_II_radius_0.1_0.1}]{\includegraphics[scale=0.24]{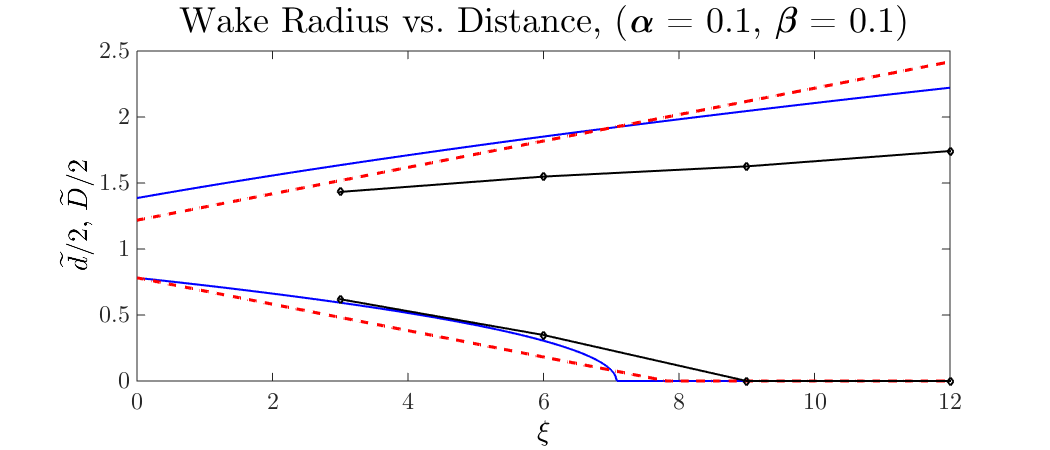}}
    \subfloat[\label{fig:comp_simulation_II_velocity_0.1_0.1}]{\includegraphics[scale=0.24]{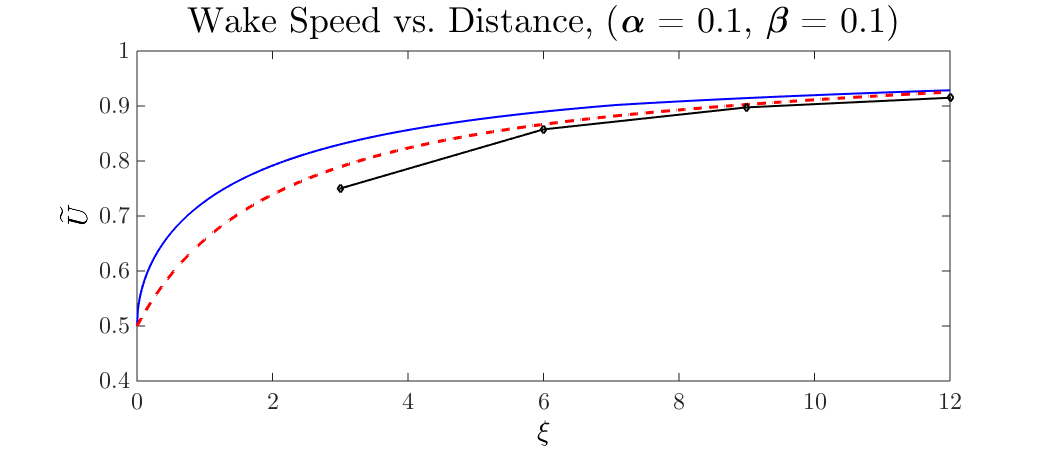}}\\
    \subfloat[\label{fig:comp_simulation_II_radius_0.05_0.05}]{\includegraphics[scale=0.24]{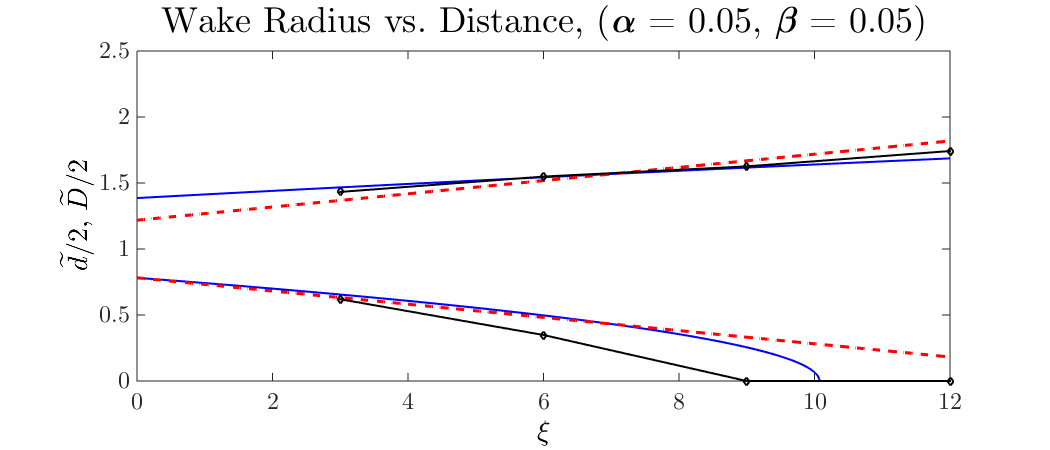}}
    \subfloat[\label{fig:comp_simulation_II_velocity_0.05_0.05}]{\includegraphics[scale=0.24]{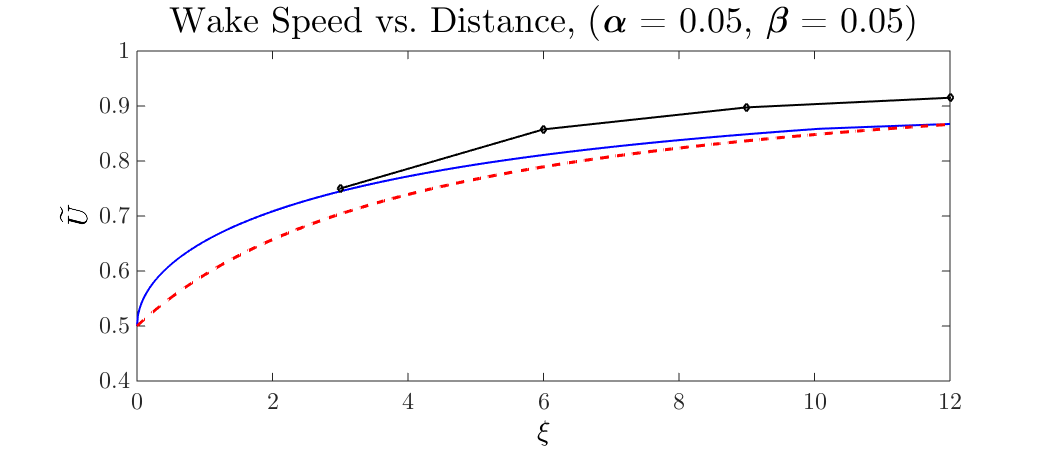}}\\  
    \subfloat[\label{fig:comp_simulation_II_radius_0.05_0.1}]{\includegraphics[scale=0.24]{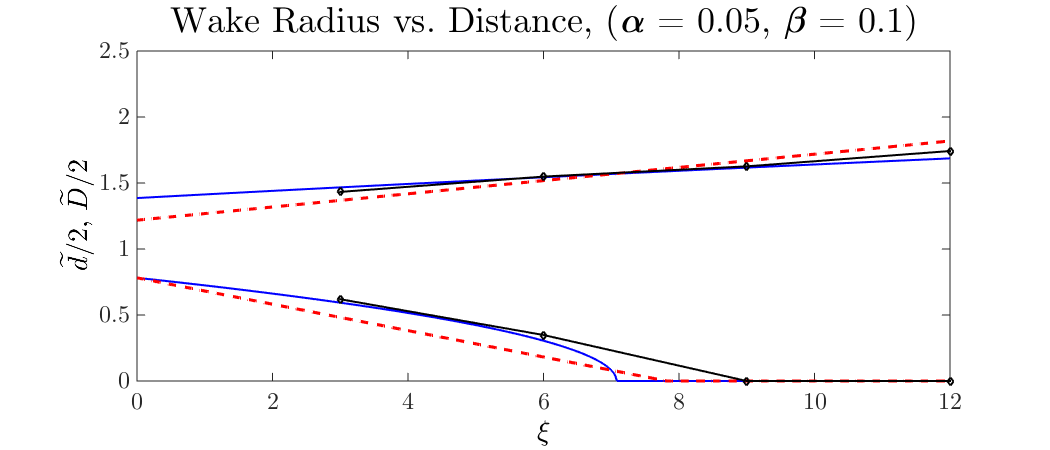}}
    \subfloat[\label{fig:comp_simulation_II_velocity_0.05_0.1}]{\includegraphics[scale=0.24]{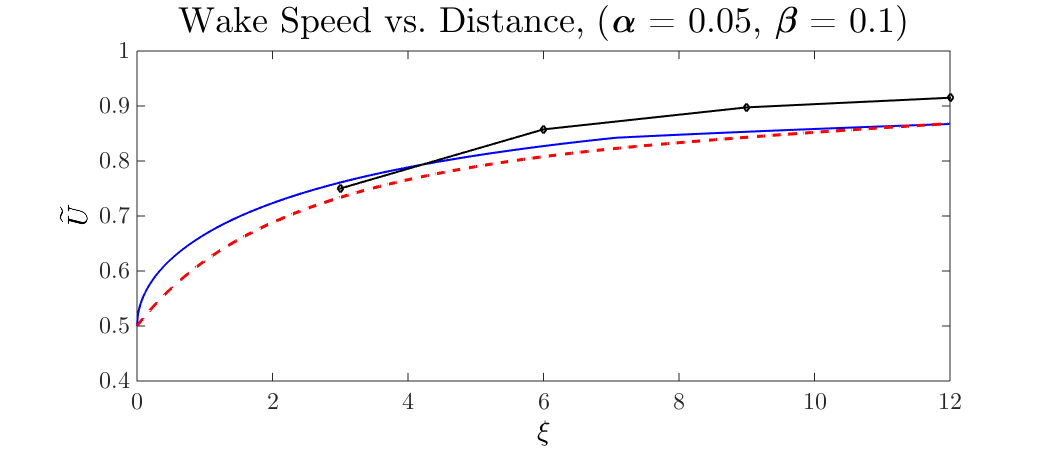}}\\
    \subfloat[\label{fig:comp_simulation_II_radius_0.0414_0.0872}]{\includegraphics[scale=0.24]{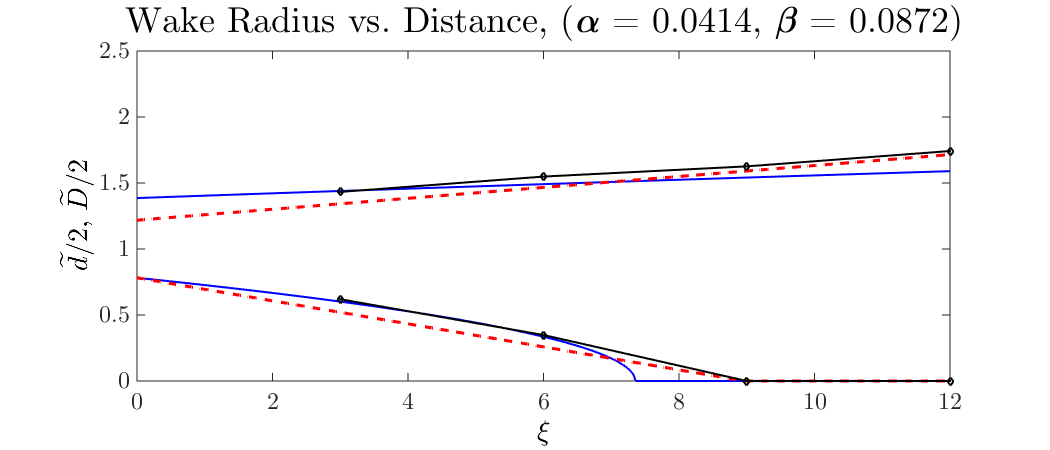}}
    \subfloat[\label{fig:comp_simulation_II_velocity_0.0414_0.0872}]{\includegraphics[scale=0.24]{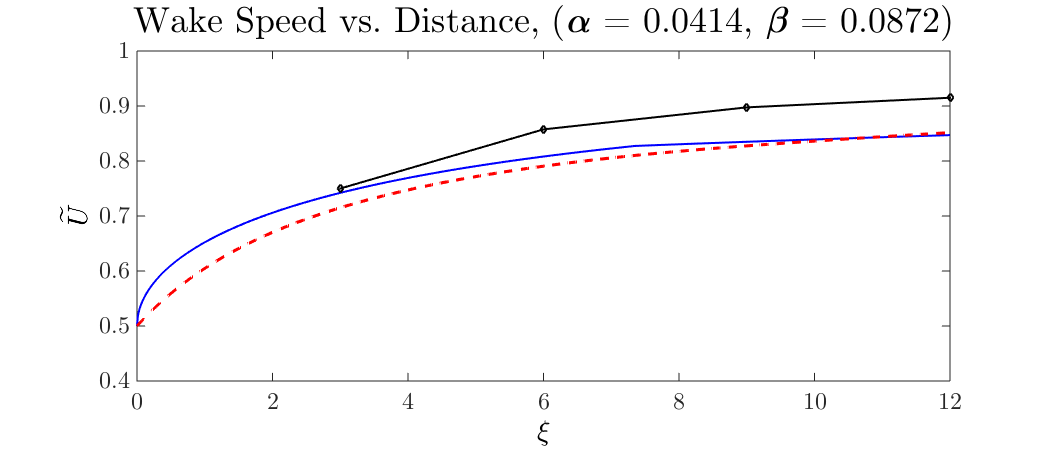}}   
    \caption{Variation of (\textit{left}) dimensionless wake radii and (\textit{right}) dimensionless wake velocity versus dimensionless distance downstream of the rotor for the CW model (dashed line), CMW model (solid line), and TI CFD model of Haas and Meyers \cite{haas-meyers} (line with markers). Various entrainment constant pairs were used in the CW and CMW models: (a,b) $\alpha=0.1$, $\beta=0.1$, (c,d) $\alpha=0.05$, $\beta=0.05$, (e,f) $\alpha=0.05$, $\beta=0.1$, and (g,h) $\alpha=0.0414$, $\beta=0.0872$.}
    \label{fig:comp_simulation_II}
\end{figure}

The next comparison that was made was with the CFD results of Haas and Meyers \cite{haas-meyers}. In particular, the analytical model was examined against their turbulent inflow results (as discussed previously, the Haas and Meyers study observed the kite system under two inflow conditions, namely laminar inflow (LI) and turbulent inflow (TI)\footnote{The turbulence was simulated using the \textit{tugen library} \cite{gilling2009tugen}.}). In this study, the kite (a wing employing an SD7032 airfoil) was flown at a constant angular velocity, $\Omega=0.369$ rad/s, in a circular path, $R=155.77$ m. Here, the uniform wind speed was $U_\infty=10$ m/s, which was assumed to be spatially uniform and flowing normal to the plane of rotation. The full list of parameters can be found in Table \ref{tab:key_parameters_Haas_Meyers}. This is the list of parameters used to simulate the analytical model for this comparison. As for the CFD simulation, the effects of the kite on the flow were simulated by applying a body force, whereupon lift was computed using actuator-line technique\footnote{This is a technique originally developed for conventional wind turbines (see Ref. \cite{sorensen2002numerical} for details).} while drag was neglected. Some details on the CFD simulation set-up can be found in the bottom half of Table \ref{tab:key_parameters_Haas_Meyers}.
\\ \\
The pairs of entrainment constants used to examine the comparisons with the TI simulation results can be found in Table \ref{tab:test_cases_simulation_II}. Here also, $\beta$ obtained from curve-fitting to CFD results is quite larger (over two times) than $\alpha$, indicating, once again, that the wake expands faster towards the core region or flow centreline. The area-averaged induction factor to be used in this study is also calculated as $a= 0.25$ by inspecting Ref. \cite[Figure 5]{haas-meyers}.
\\ \\
Figure \ref{fig:comp_simulation_II} shows the results of these comparisons. Here also, the results from the CW and CMW models are represented by dashed and solid lines, respectively, and CFD results are shown by the line with markers. As seen from the wake expansion figures, according to CFD results, both inner and outer radii expand quite linearly --- note that these results are for $\xi \ge 3$ in contrast to those in Ref. \cite{kheiriet2019} which were given for $\xi \ge 0.5$. Except for the outer radius in Figure \ref{fig:comp_simulation_II} \subref{fig:comp_simulation_II_radius_0.1_0.1}, analytical models are a good predictor of the expansion of the wake radii with the CMW model results to be slightly superior. 
The asymptotic shape of the inner radius, however, causes the difference between analytical and CFD wake radii to increase for further distances. Despite this, it can be seen that the analytical models, particularly the CW model, predict the critical point reasonably well. Looking at the wake speed curves, shown on the right column of Figure \ref{fig:comp_simulation_II}, 
it can be seen that the curves for the analytical models converge as the downstream distance increases. Generally, the analytical models underestimate the wake velocity; nevertheless, the rate of wake velocity recovery for both analytical models can be seen to be in a very good agreement with the CFD model, especially for large values of $\xi$. For example, from Figure \ref{fig:comp_simulation_II} \subref{fig:comp_simulation_II_velocity_0.1_0.1}, the maximum relative error between analytical and CFD wake velocities reduces from 11\% to 2\% as $\xi$ is increased from 3 to 12.
\begin{table}[!t]
    \centering
    \caption{Key parameters of the kite system of Haas and Meyers \cite{haas-meyers} and their CFD simulation set-up.}
    \label{tab:key_parameters_Haas_Meyers}
    \resizebox{\textwidth}{!}{
    \begin{tabular}{|c|c|c|c|c|c|c|c|c|}
       \hline
       \multicolumn{9}{|c|}{\cellcolor[gray]{0.9}\textbf{Kite System Parameters}}\\
       \hline
       $\boldsymbol{b\textbf{\;(m)}}$  & $\boldsymbol{A_k}$ \textbf{(m}$\boldsymbol{{}^2}$\textbf{)} & $\boldsymbol{R}$ \textbf{(m)} & $\boldsymbol{D_r}$ \textbf{(m)} & $\boldsymbol{d_r}$ \textbf{(m)} & $\boldsymbol{\Omega}$ \textbf{(rad/s)} & $\boldsymbol{U_\infty}$ \textbf{(m/s)} & $\boldsymbol{Re}$ & \textbf{Airfoil}
       \\
       \hline
        68 & 576 & 155.77 & 379.54 & 243.54 & 0.369 & 10 & $30 \times 10^6$ & SD7032 \\
        \hline \hline
        \multicolumn{9}{|c|}{\cellcolor[gray]{0.9} \textbf{CFD Parameters}}\\
        \hline
        \multicolumn{3}{|c|}{\textbf{Domain size (W}$\boldsymbol{\times}$\textbf{H}$\boldsymbol{\times}$\textbf{L)} \textbf{(km)}} & \textbf{Blockage ratio (\%)} & \multicolumn{2}{|c|}{\textbf{Grid resolution/count}} & \multicolumn{2}{|c|}{\textbf{Solver type}} & \textbf{Software} \\
        \hline
        \multicolumn{3}{|c|}{$0.91\times 0.91 \times 3.65$} & 8 & \multicolumn{2}{|c|}{$640\times 160 \times 320$} & \multicolumn{2}{|c|}{LES} & SP-Wind \\
        \hline
    \end{tabular}}
\end{table}
\begin{table}[!t]
    \centering
    \caption{Test Cases for Simulation II}
    \label{tab:test_cases_simulation_II}
    \begin{tabular}{|c|c|c|}
       \hline
       \cellcolor[gray]{0.9} & \multicolumn{2}{|c|}{\cellcolor[gray]{0.9}\textbf{Entrainment Constant}} \\
       \cline{2-3}
        \multirow{-2}{*}{\cellcolor[gray]{0.9}\textbf{Case Number}}  & $\boldsymbol{\alpha}$ & $\boldsymbol{\beta}$ \\
        \hline
        \cellcolor[gray]{0.9} \textbf{1} & 0.1 & 0.1 \\
        \hline
        \cellcolor[gray]{0.9} \textbf{2} & 0.05 & 0.05 \\
        \hline
        \cellcolor[gray]{0.9} \textbf{3} & 0.05 & 0.1 \\
        \hline
        \cellcolor[gray]{0.9} \textbf{4} & 0.0414 & 0.0872 \\
        \hline
    \end{tabular}
\end{table}
\subsection{Simulation III: Comparison with Haas and Meyers Results for Laminar Inflow}
\label{subsec:simulation_III_comp}
\begin{figure}[!t]
    \centering
    \subfloat[\label{fig:comp_simulation_III_radius_0.05_0.05}]{\includegraphics[scale=0.24]{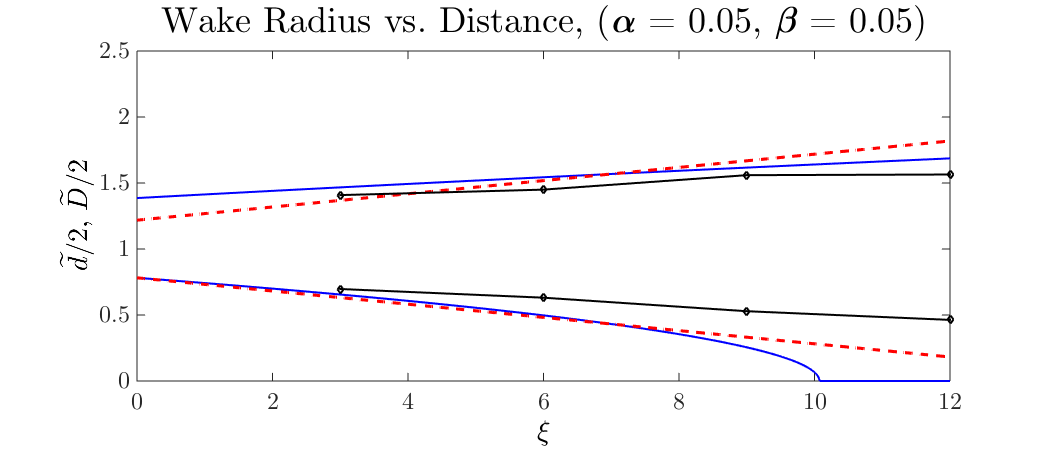}}
    \subfloat[\label{fig:comp_simulation_III_velocity_0.05_0.05}]{\includegraphics[scale=0.24]{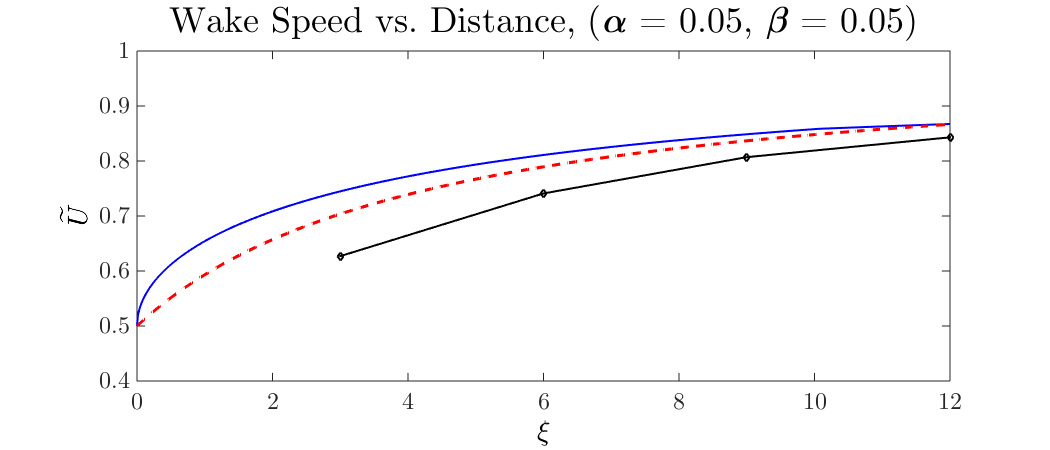}}\\
    \subfloat[\label{fig:comp_simulation_III_radius_0.025_0.025}]{\includegraphics[scale=0.24]{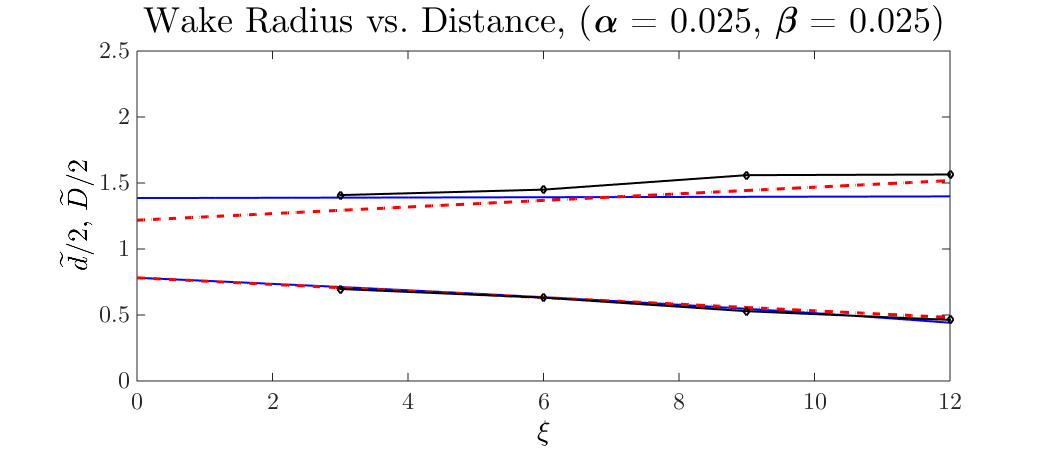}}
    \subfloat[\label{fig:comp_simulation_III_velocity_0.025_0.025}]{\includegraphics[scale=0.24]{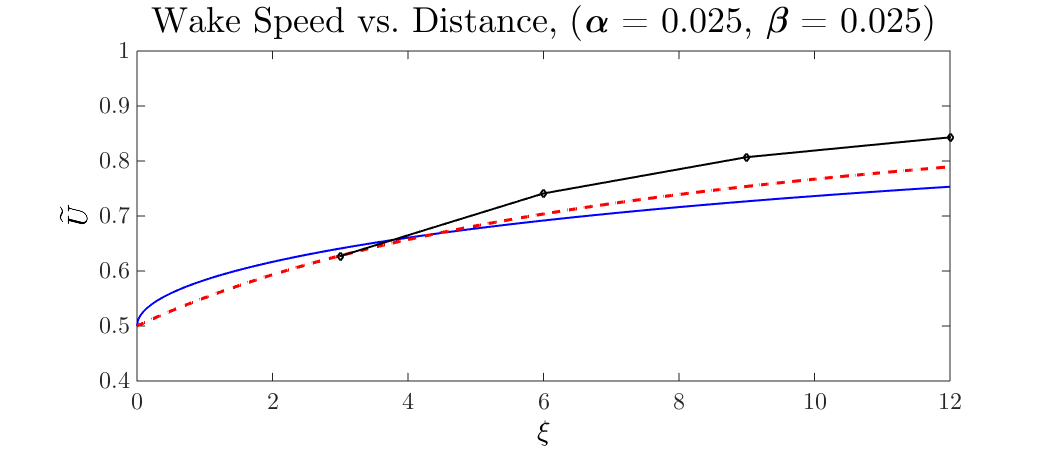}}\\  
    \subfloat[\label{fig:comp_simulation_III_radius_0.05_0.025}]{\includegraphics[scale=0.24]{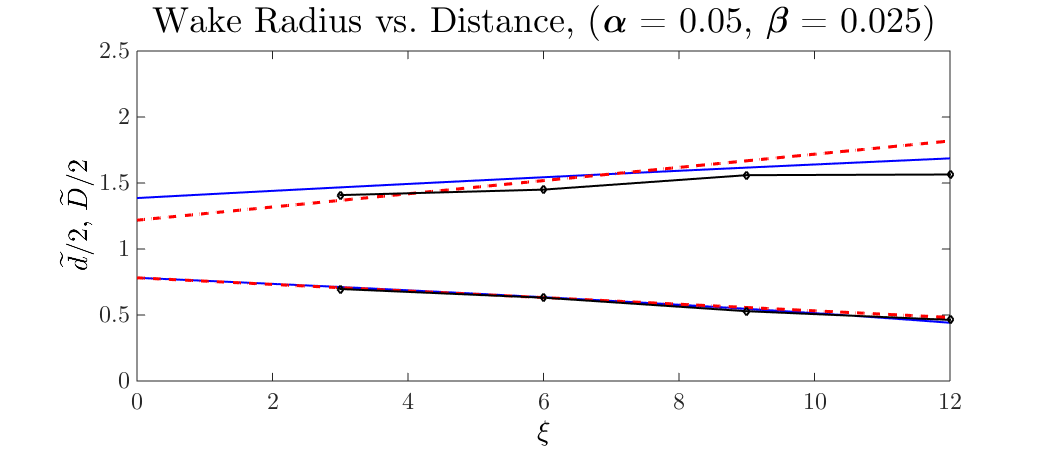}}
    \subfloat[\label{fig:comp_simulation_III_velocity_0.05_0.025}]{\includegraphics[scale=0.24]{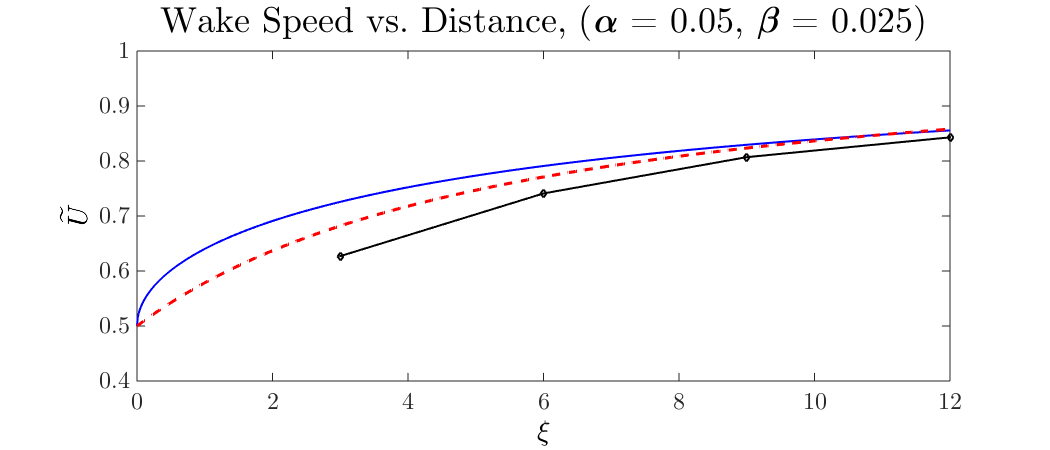}}\\
    \subfloat[\label{fig:comp_simulation_III_radius_0.0282_0.0269}]{\includegraphics[scale=0.24]{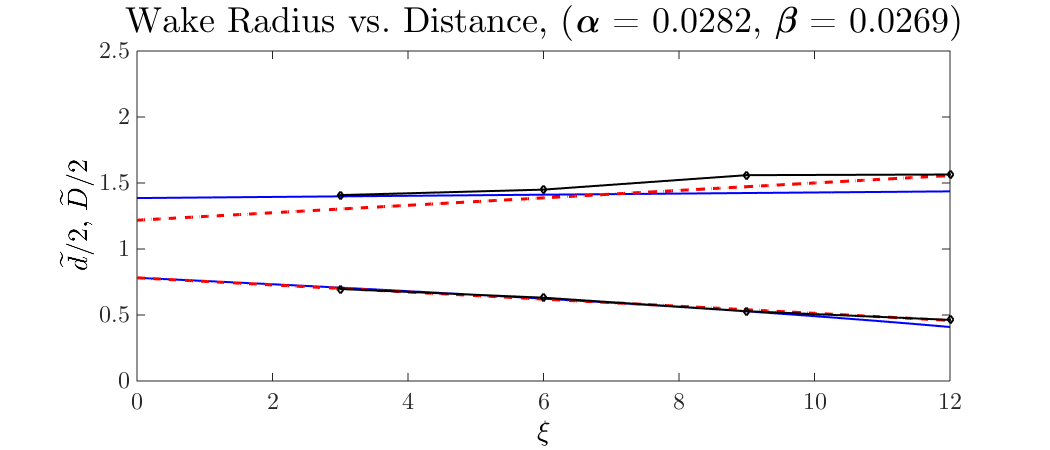}}
    \subfloat[\label{fig:comp_simulation_III_velocity_0.0282_0.0269}]{\includegraphics[scale=0.24]{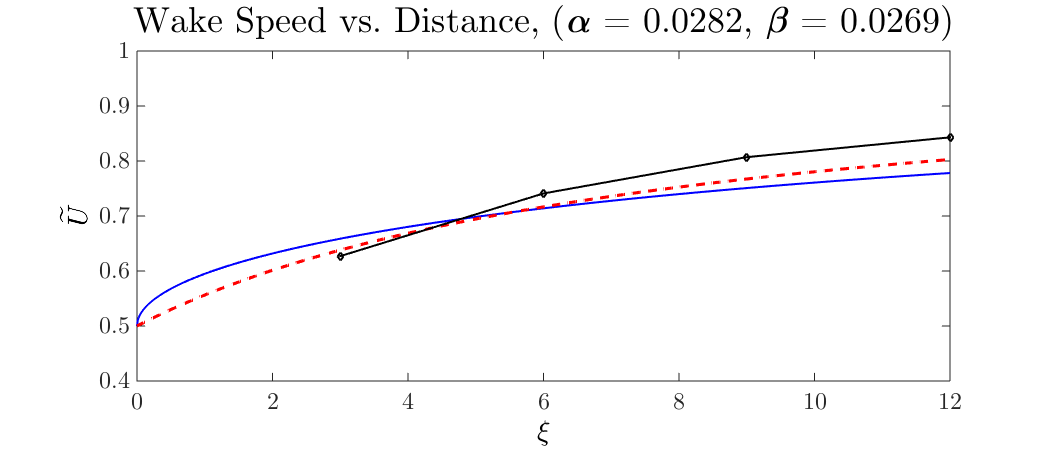}}   
    \caption{Variation of (\textit{left}) dimensionless wake radii and (\textit{right}) dimensionless wake velocity versus dimensionless distance downstream of the rotor for the CW model (dashed line), CMW model (solid line), and LI CFD model of Haas and Meyers \cite{haas-meyers} (line with markers). Various entrainment constant pairs were used in the CW and CMW models: (a,b) $\alpha=0.05$, $\beta=0.05$, (c,d) $\alpha=0.025$, $\beta=0.025$, (e,f) $\alpha=0.05$, $\beta=0.025$, and (g,h) $\alpha=0.0282$, $\beta=0.0269$.}
    \label{fig:comp_simulation_III}
\end{figure}

The third and final comparison is made against Haas and Meyers CFD results for laminar inflow (LI). The system parameters for the analytical models for this comparison were the same as those of Simulation II, given in Table \ref{tab:key_parameters_Haas_Meyers}. As discussed earlier in this section, the pairs of entrainment constants used are, however, different, and they are listed in Table \ref{tab:test_cases_simulation_III}. Interestingly, here, $\alpha$ and $\beta$ obtained by curve-fitting to CFD results are only slightly different from each other, indicating that the wake expands almost at the same rate towards the inner and outer flow regions. This is, however, different from what was observed in Sections \ref{subsec:simulation_I_comp} and \ref{subsec:simulation_II_comp}, where the wake expansion towards the inner region occurred at a higher rate. The same induction factor $a=0.25$ was used in the analytical models calculations according to \cite[Figure 5]{haas-meyers}.
\\ \\
Figure \ref{fig:comp_simulation_III} shows the comparisons for Simulation III. Here also, the CFD wake radii expand quite linearly but much more gradually compared to those for the TI simulation. This is most likely because with turbulent inflow, wake flow mixing intensifies, which results in a faster flow recovery. As seen from the figure, the analytical inner and outer radii match very well those of CFD (with Case 1 being the outlier). 
For the wake speed, depending on the values of $\alpha$ and $\beta$, analytical models may overestimate (Figures \ref{fig:comp_simulation_III} (\subref*{fig:comp_simulation_III_velocity_0.05_0.05},\subref*{fig:comp_simulation_III_velocity_0.05_0.025})) or underestimate (Figures \ref{fig:comp_simulation_III} (\subref*{fig:comp_simulation_III_velocity_0.025_0.025},\subref*{fig:comp_simulation_III_velocity_0.0282_0.0269})) with respect to CFD results. For the former, the gap between analytical and computational results narrows as $\xi$ is increased whereas for the latter the gap widens. Interestingly, the CW model results are in a better numerical agreement with the CFD results. 


\begin{table}[!t]
    \centering
    \caption{Test Cases for Simulation III}
    \label{tab:test_cases_simulation_III}
    \begin{tabular}{|c|c|c|}
       \hline
       \cellcolor[gray]{0.9} & \multicolumn{2}{|c|}{\cellcolor[gray]{0.9}\textbf{Entrainment Constant}} \\
       \cline{2-3}
        \multirow{-2}{*}{\cellcolor[gray]{0.9}\textbf{Case Number}}  & $\boldsymbol{\alpha}$ & $\boldsymbol{\beta}$ \\
        \hline
        \cellcolor[gray]{0.9} \textbf{1} & 0.05 & 0.05 \\
        \hline
        \cellcolor[gray]{0.9} \textbf{2} & 0.025 & 0.025 \\
        \hline
        \cellcolor[gray]{0.9} \textbf{3} & 0.05 & 0.025 \\
        \hline
        \cellcolor[gray]{0.9} \textbf{4} & 0.0282 & 0.0269 \\
        \hline
    \end{tabular}
\end{table}

\clearpage
\section{Discussion and concluding remarks}
\label{sec:discussion}
Some general observations may be made based on the plots shown in Figures \ref{fig:comp_simulation_I} to \ref{fig:comp_simulation_III}. First, we can see that the wake velocity recovery occurs faster for the CMW model than the CW model, and hence, the wake is generally narrower in spite of the fact that the CMW model gets a ``head start" (in that it has a higher outer diameter at $\xi=0$). Second, we notice that, in most cases, the error between the analytical and CFD curves tends to decrease as $\xi$ increases. As pointed out in Bastankhah and Port\'e-Agel \cite{Bastankhah2014} for conventional wind turbines, one of the reasons for this is our assumption of axisymmetry: the effects of the kite on the rotation of the flow start to disappear far downstream of the rotor plane. The larger error before this ``far wake" region is not of great concern to us, since eventually when distances between kites are studied, it is the far downstream distances that will be of practical interest. Third, the analytical results and the level of agreement between them and CFD results are heavily dependent on the values of entrainment constants. Despite a wide range for entrainment constants, results from CW and CMW models showed little difference in both their predicted wake shape distribution and velocity distribution, and they were in fairly good agreement with the CFD results.
\\ \\
Several reasons may be behind the small discrepancy observed between analytical and computational results. As also indicated by Bastankhah and Port\'e-Agel, the assumption of a top-hat (or uniform) flow velocity distribution in the wake may be one of the main reasons. Computational studies of Kheiri et al. \cite{kheiriet2019} and Haas and Meyers \cite{haas-meyers} show that the velocity distribution in the wake is non-uniform, and it is similar to the Gaussian distribution. Another reason may be that the flow velocity in the core region was assumed to remain unchanged at the freestream velocity while CFD results showed that it may increase/decrease as a function of axial distance. Other contributing factors to the difference between analytical and CFD results may be the assumption of a negligible initial wake expansion length and that of no flow rotation. In the case of the CW model, the fact that only mass conservation is considered in the development of the model may also be an important factor.  
\\ \\
As discussed previously, entrainment constants are generally dependent on the type of terrain, kite altitude and, according to some studies on conventional wind turbines like Ref. \citep{Andersen2014}, on the thrust coefficient (and hence also the induction factor) of the wind energy extracting device. On the other hand, previous studies on conventional wind turbines, e.g. \cite{Kaldellis2021,crespo1999survey}, have shown that the atmospheric boundary layer may greatly affect turbulent mixing and thus wake development. Considering the fact that neither the CFD work of Kheiri et al. nor that of Haas and Meyers considered the effects of the atmospheric boundary layer, the strong demand for fresh CFD studies that consider such effects is felt. Such CFD studies, however, would not obviate the need for field tests or, alternatively, wind tunnel tests. 

\section*{Acknowledgments}
The support for this research project by the Natural Sciences and Engineering Research Council of Canada (NSERC) is gratefully acknowledged.

\clearpage
\bibliographystyle{elsarticle-num}
\bibliography{References.bib}

\clearpage
\newpage

\appendix
\section{Detailed derivation of the momentum equation}
\label{sec:appendix_A}
From continuity, we can write
\begin{equation}
    \int_{A_i}\rho U_\infty \ dA=\int_{A}\rho U \ dA+\int_{A_c}\rho U_\infty \ dA + \int_{A_l}\rho \boldsymbol{U} \cdot d \boldsymbol{S},
    \label{eq:App_A_continuity}
\end{equation}
where $A_i$, $A$, $A_c$, and $A_l$ are the areas of the control volume inlet, annular portion of the outlet, core portion of the outlet, and lateral/curved surface, respectively.
\\ \\
Thus, in the above continuity equation, the term on the l.h.s. is the mass flow going into the control volume through the inlet while the three terms on the r.h.s. represent the mass flow leaving the control volume through the annular portion of the outlet, core portion of the outlet, and the lateral surface, respectively. 
\\ \\
One can easily multiply a constant value (e.g. $U_\infty$) to the two sides of equation (\ref{eq:App_A_continuity}): 
\begin{equation}
    \int_{A_i}\rho U_\infty^2 \ dA=\int_{A}\rho U U_\infty \ dA+\int_{A_c}\rho U_\infty^2 \ dA + \int_{A_l}\rho U_\infty (\boldsymbol{U} \cdot d \boldsymbol{S}).
    \label{eq:App_A_continuity_modified}
\end{equation}
Now, let us consider the momentum equation (\ref{eq7-1})
\begin{equation}
	-T=-\int_{A_i}\rho U_\infty^2\;dA+\int_{A}\rho U^2\;dA+\int_{A_c}\rho U_\infty^2\;dA + \int_{A_l}\rho (\boldsymbol{U} \cdot \boldsymbol{\widehat{e}_x}) (\boldsymbol{U} \cdot d \boldsymbol{S}),
	\label{eq:App_A_momentum_first}
\end{equation}
where $\boldsymbol{\widehat{e}_x}$ is the unit vector in the $x$-direction.
\\ \\
It is argued that for the flow to be able to leave the control volume through the lateral surface, there should be a radial velocity component. This is evident from the continuity equation (\ref{eq:App_A_continuity}) where $d \boldsymbol{S} = (d S) \boldsymbol{\widehat{e}_r}$, with $\boldsymbol{\widehat{e}_r}$ being the unit vector in the radial direction. To simplify the formulation, it is assumed that this radial velocity component is very small compared to the axial component and that it exists only within an infinitesimally small radial distance close to the lateral surface, outside of which the flow velocity is parallel to the $x$-axis and equal to $U_\infty$. Thus, equation (\ref{eq:App_A_momentum_first}) may be simplified to
\begin{equation}
	-T=-\int_{A_i}\rho U_\infty^2\;dA+\int_{A}\rho U^2\;dA+\int_{A_c}\rho U_\infty^2\;dA + \int_{A_l}\rho U_\infty (\boldsymbol{U} \cdot d \boldsymbol{S}).
	\label{eq:App_A_momentum_second}
\end{equation}
Now, we replace the first term on the r.h.s. of equation (\ref{eq:App_A_momentum_second}) with the expression obtained on the r.h.s. of equation (\ref{eq:App_A_continuity_modified}):
\begin{align}
	-T= & -\int_{A}\rho U U_\infty \ dA -\int_{A_c}\rho U_\infty^2 \ dA - \int_{A_l}\rho U_\infty (\boldsymbol{U} \cdot d \boldsymbol{S}) \nonumber \\
	&+\int_{A}\rho U^2\;dA+\int_{A_c}\rho U_\infty^2\;dA + \int_{A_l}\rho U_\infty (\boldsymbol{U} \cdot d \boldsymbol{S}),
	\label{eq:App_A_momentum_third}
\end{align}
which is finally simplified to
\begin{align}
	T= \int_{A}\rho U (U_\infty - U) \ dA,
	\label{eq:App_A_momentum_final}
\end{align}
that is exactly the same as equation (\ref{eq8}).

\end{document}